\documentclass[12pt]{article}

\usepackage{latexsym,amsmath,amssymb,amsthm,amsfonts,graphicx,natbib,breqn,epsfig,bm,authblk,url,tikz, hyperref, subcaption, algorithm, algorithmic}

\PassOptionsToPackage{pagebackref}{hyperref}

\usepackage{float} 
\usepackage[nameinlink]{cleveref}
\Crefname{figure}{Fig.}{Figures}

\usepackage{booktabs,comment} 
\usepackage[margin=1cm]{caption} 
\usepackage{xcolor}
\usepackage{soul}

\floatstyle{plain}
\newfloat{Algorithm}{thp}{lop}
\floatname{Algorithm}{Algorithm}

\begin{document}

\title{Summary Statistics of Large-scale Model Outputs for Observation-corrected Outputs}

\date{\empty}

\author{Atlanta Chakraborty, Julie Bessac}
\affil{National Renewable Energy Laboratory, 15013 Denver W Pkwy, Golden, CO 80401}

\maketitle

\vspace{-0.3in}

\begin{abstract}
Physics-based models capture broad spatial and temporal dynamics, but often suffer from biases and numerical approximations, while observations capture localized variability but are sparse. Integrating these complementary data modalities is important to improving the accuracy and reliability of model outputs. Meanwhile, physics-based models typically generate large outputs that are challenging to manipulate.
In this paper, we propose Sig-PCA, a space-time framework that integrates summary statistics from model outputs with localized observations via a neural network (NN). By leveraging reduced-order representations from physics-based models and integrating them with
observational data, our approach corrects model outputs,  while allowing to work with dimensionally-reduced quantities hence with smaller NNs. This framework highlights the synergy between observational data and statistical summaries of model outputs, and effectively combines multisource data by preserving essential statistical information. We demonstrate our approach on two datasets (surface temperature and surface wind) with different statistical properties and different ratios of model to observational data.  Our method corrects model outputs to align closely with the observational data, specifically enabling to correct probability distributions and space-time correlation structures.
\end{abstract}

\section{ Introduction}

Physics-based models, such as atmospheric models, typically simulate physical systems and are driven by mathematical representations of the system, which require computational treatment to output simulations. While they provide consistent outputs across space and time, they are often subject to model biases arising from approximations in physical processes, numerical approximations, and unresolved processes from parameterization schemes \citep{danforth2007estimating,palmer2009,christensen2022}. Their high-dimensional outputs create computational difficulties in movement, storage, analysis, and visualization, raising the needs for data compression \cite{jian2024cliz}, surrogates and dedicated analysis tools such as in-situ treatment \citep{dutta2021situ}. 
In contrast, measurement observations capture localized, model-independent information, but are typically sparse in space and time and subject to measurement errors. 
Observational data are essential for validating and improving model outputs, improving their representations, and understanding local phenomena, especially in complex environments, where topography or local environments significantly influence patterns. 

To improve model outputs accuracy and realism, techniques to directly correct them have been proposed in various settings such as post-processing of forecasts \citep{glahn1972,raftery2005,vannitsem2021statistical} or bias correction \citep{maraun2019statistical}. 
Bias correction aims at matching the statistical properties of physics-based model outputs to the observation ones. Statistical methods have been extensively used in that field. 
Quantile-matching and its extensions provide an empirical transformation from the model output distribution to the observational one \citep{maraun2016,cannon2018}. 
Similarly, statistical parametric models have been built to correct physics-based model outputs, leveraging various statistical features and interactions across data sources or variables, \citep{bessac+ca18,majumder2024spatiotemporaldensitycorrectionmultivariate}.  
However, including spatial, multivariate or multi-source information remains a statistical challenge, and general techniques are lacking due to limitations in statistical methods for complex dependencies and high-dimensional settings, \citep{cressie2008,salvana2020nonstationary}. 
In a similar vein, downscaling enhances the resolution of model outputs when scales are mismatching while capturing finer-scale statistical properties. Recent advances in deep learning have lead to a profusion of literature in downscaling, from generative adversarial networks and their stochastic extension \citep{stengel2020adversarial,annau+cm23,daust+m24} to normalizing flows \citep{groenke+mm21,mcdonald+tl22,bailie2024quantile}. 
However, most of these methods require the entire gridded model outputs to learn corrective models. 

To address the challenge of handling large datasets, unsupervised dimension reduction techniques such as principal component analysis (PCA) \citep{von2002statistical,hannachi2007empirical} and its variants, such as non-linear ones \citep{monahan00}, have been widely used to extract dominant spatial and temporal modes from large fields.  
Deep learning methods have also been employed in dimension reduction, such as variational autoencoders (VAE) providing a probabilistic low-dimensional embedding of complex high-dimensional fields, \citep{kingma2022autoencoding}. However, their interpretation remains an active area of research, e.g. in disentangling and enhancing the interpretability of VAE latent space for scientific data \citep{ganguli2024enhancing}. 
Meanwhile, implicit neural representations (INR) offer a complementary approach by encoding complex spatial and temporal fields as continuous functions parameterized by NN, as exemplified in \citep{qayyum+etal24} in a wind velocity super-resolution context.  
In addition to dimension reduction, data compression such as error-bounded lossy compression \citep{di2024survey} has gained significant traction for storage and movement of scientific data, including interests in atmospheric model outputs and their statistics \citep{poppick2020statistical,underwood2022b,qian+etal23}. In the meantime, specific compression techniques for atmospheric data have been developed by \cite{klower2021compressing} and \cite{huang2022compressing} based on information theory and NN representations respectively.      
In the statistical context, \cite{Guinness02012018} have proposed a statistical compression and decompression method that stores Fourier coefficients as essential summary statistics, along with a decompression statistical model describing the conditional distribution of the full dataset based on these summaries.  

In this paper, we propose a novel approach, called Sig-PCA, to integrate space-time reduced representations of  numerical model outputs with observational data in order to correct model outputs. 
To the best of our knowledge, this is the first attempt to bias-correct physics-based model outputs from reduced representations of the former. 
Specifically, we propose a NN-based approach that inputs space-time summary statistics rather than the full-resolution of model outputs, and generates outputs matching measurement data, hence correcting the model outputs in space and time. By processing low-dimensional representations derived from model outputs with observations, we enable a computationally efficient integration of multisource data, while capturing meaningful patterns (probability distribution and space-time correlation structures) without handling the entire model output complexity. 
Path signatures are used to generate summary statistics in an unsupervised fashion, as opposed to deep-learning based dimension reduction models requiring training, hence ensuring more tractable computations and interpretability.  
Moreover, the proposed NN are inputted with deep kriging basis functions \cite{chen2022deepkrigingspatiallydependentdeep}, allowing to extend the learned observation-based correction to a spatial neighborhood of the considered measurement stations. 
Finally, the method captures the spatiotemporal non-stationarity of the data via the summary statistics and deep-kriging.

The paper is outlined as follows: 
Section \ref{sec:methods} details the proposed methodology, beginning with an introduction to path signatures, followed by the presentation of the proposed NN-based corrective method, Sig-PCA. Section \ref{sec:data} presents the datasets (surface temperature and surface wind speed) comprised of model outputs and observations. Each data application presents a different type of model data (reanalysis and numerical weather prediction), a different ratio of model to observation data, and different statistical properties as temperature exhibits smoother spatiotemporal behaviors than surface wind speed. 
Sections \ref{sec:temperature_results} and \ref{sec:wind_results} demonstrate the application of our approach to these datasets and provide statistical validation. Finally, Section \ref{sec:conclu} concludes the paper, summarizing key findings and potential future directions for research.

\section{Sig-PCA Proposed Method}\label{sec:methods}

First, we introduce the concept of path signatures, which serves as an unsupervised dimension reduction technique for time series data. Second, we present the proposed method, Sig-PCA (Signature Principal Component Analysis), which trains a NN on summary statistics of model outputs to learn localized observational features. Sig-PCA enables to extract compact representations of model outputs, and enhances their spatiotemporal reconstruction and correction with observations.

\subsection{Path Signatures}
Path signatures capture the temporal structure of a time series object via a collection of computed iterative integrals, summarizing a complex, high-dimensional process, while retaining important information about the system's evolution, particularly in the presence of intricate temporal dependencies or non-linear behaviors, \citep{chevyrev+k16}.  
\cite{lyons+s05} introduced an approach to sound compression using the signature method, which outperforms traditional Fourier and wavelet transforms by accounting for non-linear dependencies. \cite{dyer+cs24} applied path signatures as sufficient summary statistics in approximate Bayesian computation \citep{abc} to address the sequential nature of time series, demonstrating effective Bayesian parameter inference for simulators. \cite{morrill+fkl21} introduced a domain-agnostic generalized signature method for extracting features of multivariate time series, unifying different variations of the signature method into a cohesive framework. We leverage their approach in the following.

\subsection{Computing Path Signatures}\label{sec:pathsig}
Consider a multidimensional time series \( x \), as a \( D \times T \) matrix, where \( x_{i,t} \) denotes the value at the \( i \)-th location and the \( t \)-th time instant. Here, \( i = 1, \dots, D \), and \( t \) spans \( T \) discretized time intervals between \( t_0 \) and \( t_1 \). In our examples, we have an equal number of time samples per location. The depth-N signature of the time series is defined as: 
\begin{align}
    Sig^N(x) &= \left( S(x)^{i}_{i=1, \cdots,D}, \ S(x)^{i,j}_{i,j=1, \cdots,D},\  \cdots, \  S(x)^{i_1,i_2, \cdots, i_N}_{i_1,i_2,\  \cdots, i_N=1, \cdots,D} \right)
\end{align}
where 
$$S(x)^{i_1,i_2, \cdots, i_d}_{i_1,i_2,\  \cdots, i_d=1, \cdots,D}= \int_{t_0\leq u_1 \leq u_2 \leq \cdots \leq u_d \leq t_1} dx_{i_1, u_1}dx_{i_2, u_2}\cdots dx_{i_d, u_d}.$$
The depth-1 term of the signature is $(S(x)^1, \cdots, S(x)^D)= ( \int_{t_0 \leq u \leq t_1} dx_{1,u}, \cdots, \int_{t_0 \leq u \leq t_1} dx_{D,u})= (x_{1,t_1}-x_{1,t_0}, \cdots, x_{D,t_1}-x_{D,t_0})$ denotes the displacement over time for the multidimensional path $(x_{1,t}, \cdots, x_{D,t})$ within $[t_0, t_1].$ It takes $\mathcal{O}(DT)$ to compute depth-1 signatures. The depth-2 terms of the signature, 
\begin{align}
    &(S(x)^{1,1}, \cdots, S(x)^{i,j}, \cdots, S(x)^{D,D}) = \notag \\
    &\quad \left( \int_{t_0 \leq u_1 \leq u_2 \leq t_1} dx_{1,u_1} dx_{1,u_2}, \cdots, 
    \int_{t_0 \leq u_1 \leq u_2 \leq t_1} dx_{i,u_1} dx_{j,u_2}, \cdots, \int_{t_0 \leq u_1 \leq u_2 \leq t_1} dx_{D,u_1} dx_{D,u_2} \right),
\end{align}
represents the signed area enclosed by the path traced by the $i-$th and $j-$th dimensions of the time series over $[t_0,t_1].$ Computing depth-2 naively takes $\mathcal{O}(DT^2).$ 

We adopt the generalized signature method proposed in \cite{morrill+fkl21} that combines the following existing techniques into a single framework to extract path signatures from a time series. In their work, the features of a time series, $x$, are expressed as
$\displaystyle (Sig^N \circ W \circ \phi)(x)$ where 
\begin{itemize}
    \item $\phi$ represents a basepoint and time-augmentation transformation, i.e., augmenting the time series $x$ to $(0,x)$, which is then followed by embedding the time dimension into the path as $(t, x_{1,t}, \cdots , x_{D,t})$ for $t \in [t_0, t_1].$ While the time-augmentation transformation increases the dimensionality, it provides information about the temporal structure of the time series. Adding $0$ as the basepoint transformation captures variations in the path's starting position. 
    \item $W$ denotes a windowing operation, which is used to decompose the time interval \([t_0, t_1]\). This allows us to extract features over specific time intervals. Since model outputs contain multiscale information, we employ hierarchical dyadic windows to learn the structure of the data across different levels of granularity. The dyadic windowing approach recursively divides the data into progressively smaller time windows. Initially, the entire time period \([t_0, t_1]\) is treated as a global window. This window is then split into two sub-windows: the first half, \([t_0, (t_0+t_1)/2]\) and the second half, \([(t_0+t_1)/2,t_1]\). The process continues, with each sub-window being further divided into quarters, and so on, down to a specified depth \( d_w \). Hierarchical dyadic windowing is particularly advantageous as it respects the multiscale nature of the data. While sliding and expanding windows are analogous to short-term Fourier transforms, hierarchical dyadic windows offer a structured approach to capturing multiscale patterns effectively.

For our analysis, we set the window depth to 
$d_w=3,$ meaning the data is divided into $d_w+1$ sub-windows, resulting in a total of 7 sub-windows (depth 1 (a single global window), depth 2 (two sub-windows), and depth 3 (four sub-windows)). Going beyond depth 3 becomes non-trivial, as we work with daily and hourly data, with a maximum of 24 observations per day. As the window depth increases, the sub-windows become increasingly smaller, leaving very few data points within each sub-window for computing integrals. While larger depths capture finer-scale information and preserve intricate details, the limited number of observations in extremely small sub-windows can make the computations less reliable.
\end{itemize}

\subsection{PCA-Reduced Path Signatures}
Since path signatures for a single time series at a fixed location are computed over several sub-windows, applying this to model outputs with thousands of grid locations significantly expands the signature space. To address this, we perform a PCA on the signature space to obtain a linear combination of signatures across locations, reducing dimensionality while retaining
$99.5\%$ of the variance. This transformation captures the most essential information and serves as input to the NN, as described in the next section. The algorithm is called Sig-PCA, as it integrates path signatures with PCA for efficient dimensionality reduction.

\subsection{Reconstruction and Corrective Neural Networks}\label{sec:correction}
The PCA reduced path signatures from the previous section are used as inputs for two NNs: a reconstruction network and a corrective network. 
Instead of directly learning the observations—which would require a large amount of observational data—we adopt a two-step approach: first reconstructing the model outputs and then learning the observation-driven corrections. This strategy ensures that the NN effectively leverages the available model information while remaining robust even with very few observation stations. By focusing on systematic discrepancies rather than raw observations, the NN learns structured biases in the data rather than memorizing location-specific values. This improves the model's ability to spatially interpolate across different spatial domains.

\begin{itemize}
\item{\textbf{Reconstruction Network:}} We train the network to predict a small subset of model-output gridpoints, denoted as $x\%$ of the total number of gridpoints, while using summary statistics from the full grid as inputs. The network learns to predict time series at these subset gridpoints, which are then extrapolated to recover the entire spatial grid, allowing to infer the entire grid of model outputs. The value of $x$ is selected based on practical constraints and the desired percentage root-mean square error (RMSE), illustrated in Section \ref{sec:res_recons}.

\item{\textbf{Corrective Network:}} After reconstruction of the model outputs, the PCA-reduced path signatures are used to train a second NN that learns observation-based corrections and their localized variability. The network outputs correction terms that are defined as the difference between the observations and the nearest model-output gridpoints.
\end{itemize}

To spatially interpolate across the entire grid, kriging is used to predict at unobserved locations by leveraging spatial dependence structures \citep{matheron1967kriging, cressie2015statistics}. Traditionally, kriging is applied under stationarity assumptions, but many datasets are often more complex, requiring a non-stationary covariance structure. \cite{chen2022deepkrigingspatiallydependentdeep} introduced deep kriging, which utilizes deep NNs, and radial basis functions and spatial coordinates as input to model spatial dependence. They rely on the Karhunen–Lo\'eve (KL) theorem that allows a spatial process to be expanded into a finite linear combination of orthogonal basis functions.  
Chen et al. used Wendland functions, which offers computational advantages over other radial basis functions like Gaussian and Mat\'ern, because of their compact support inducing sparsity, hence making them better suited for large-data manipulation. 
Additionally, in \cite{chen2022deepkrigingspatiallydependentdeep}, basis functions are set in a multi-resolution fashion where a sequence of grids is considered with granularity increasing by a factor 2. 
At each resolution, the DK Wendland scale/bandwidth parameter is set to 2.5 times the associated knot spacing. 
In this work, we use the above multi-resolution Wendland basis functions as inputs to our NN.

\begin{figure}
\centering
\includegraphics[width=0.85\textwidth]{ 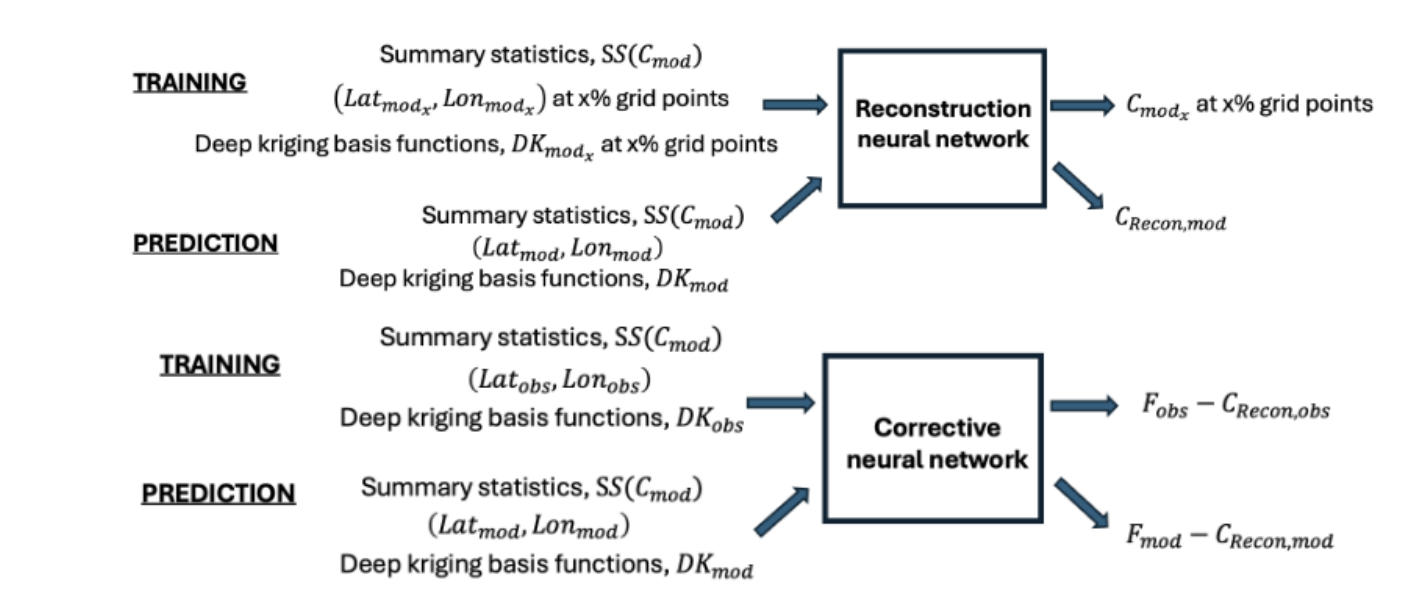}
\caption{{\bf Top} network focuses on reconstructing model data from reduced signatures features, {\bf bottom} network corrects errors between observations and gridded reconstructions. $C$ represents the model outputs, $F$ the  observations, $SS$ the path signatures summary statistics,  $DK$ the deep kriging basis functions, and $(Lat,Lon)$ the spatial coordinates. The subscript $Recon$ indicates the reconstruction process and $mod$ the model gridpoints. For each network, the top three inputs are used during the training phase, while the bottom three serve as inputs for the prediction phase.}
\label{fig:nn}
\end{figure}

Figure~\ref{fig:nn} provides an overview of the inputs and outputs of the NNs used for both reconstruction and correction. The top figure illustrates the architecture of the reconstructive NN, which aims to reconstruct the original gridded model outputs. The bottom figure represents the corrective NN, which refines the reconstructed output by learning the discrepancies between the observations and reconstructed model output. Together, these networks enhance data reconstruction and correction. We detail the specific architecture used in each result section.  
Algorithm~\ref{alg:SigPCA} outlines our proposed procedure for obtaining the observation-corrected model outputs. 

\begin{algorithm}
\caption{Sig-PCA-based reconstruction and correction algorithm}
\label{alg:SigPCA}
\begin{algorithmic}[1]
\STATE \textbf{Input:} Gridded physics-based spatiotemporal data $C_{mod}$ at $mod$ grid locations rearranged as \( (\texttt{number of samples, global time window, length(mod)}) \), observational spatiotemporal data $F_{obs}$ at $obs$ observation locations, dyadic window depth $d_w$, signature depth-N.

\STATE \textbf{Output:} Observation-corrected model outputs $C^{obs-corrected}_{mod}$ at $mod$ grid locations 

\STATE \textbf{Step 1:} \FOR{each dyadic window depth $d_w$}
    \STATE Compute path signature of $C_{mod}$ of depth-N for the window. 
\ENDFOR

\STATE \textbf{Step 2:} Perform PCA on all the path signatures to obtain further reduced representation. Store the $k$ principal components, where  $k$ is chosen to capture $99.5\%$ of the total variance, representing the desired reduced dimensionality.

\STATE \textbf{Step 3:} Use the PCA-reduced representation as input to train the reconstruction NN to learn the reconstruction, $C^{recon}_{mod_x}$ for $mod_x,$ representing $x\%$ of the total gridded locations from $mod.$

\STATE \textbf{Step 4:} Predict $C^{recon}_{mod}$ across all $mod$ locations. 

\STATE \textbf{Step 5:} Use the PCA-reduced representation as input to train the corrective NN to learn the corrections, $Corr_{obs}= F_{obs} - C^{recon}_{obs'}$ for $obs'$ locations that are closest to the $obs$ locations. 

\STATE \textbf{Step 6:} Predict the corrections at $mod$ locations as $Corr_{mod}.$

\STATE \textbf{Step 7:} At grid locations, evaluate the observation-corrected model outputs as $C^{obs-corrected}_{mod} = C^{recon}_{mod} + Corr_{mod}$.

\STATE \textbf{End}
\end{algorithmic}
\end{algorithm}

\subsection{Benchmark Reduction Method}\label{sec:eof}
Empirical orthogonal functions (EOFs) have long been a fundamental tool for decomposing spatiotemporal fields into spatial and temporal components \citep{hannachi2007empirical}, and are used for dimensionality reduction, identifying key patterns while preserving most of the variability. 
EOF analysis identifies a set of orthogonal spatial patterns along with corresponding uncorrelated time series, known as principal components (PCs), and decomposes a continuous space-time field 
$X(t,s),$ where 
$t$ represents time and 
$s$ denotes spatial position, into: 
$ \displaystyle X(t,s)= \sum_{k=1}^M f_k(t)g_k(s)$ where $M$ is the number of modes in the field, $g_k(s)$ are spatial basis functions, and $f_k(t)$ are the corresponding temporal coefficients. The truncation order, $M,$ is typically chosen based on the amount of variance retained. In our case, we select the leading EOFs that collectively capture at least $99.5\%$ of the total variance. 
 To apply the EOF consistently with Sig-PCA, we arrange the model data in a space-time structure with dimension  
\texttt{(number of samples, global time window $\times$ length(mod)),} at $mod$ grid locations.

\section{Data}\label{sec:data}
Two application datasets (surface temperature and surface wind speed) are considered to exemplify the proposed method. Each application contains a set of gridded numerical model outputs and corresponding observational ground measurements, typically sparse in space.  
The first application relies on reanalysis of surface air temperature and the second model outputs consist of numerical weather prediction (NWP) of surface wind speed.  Temperature data exhibit smoother behaviors than surface wind speed, illustrating different responses of the proposed method. 

\subsection{Surface Temperature Data}
\begin{figure}
    \centering
    
\includegraphics[width=\textwidth]{   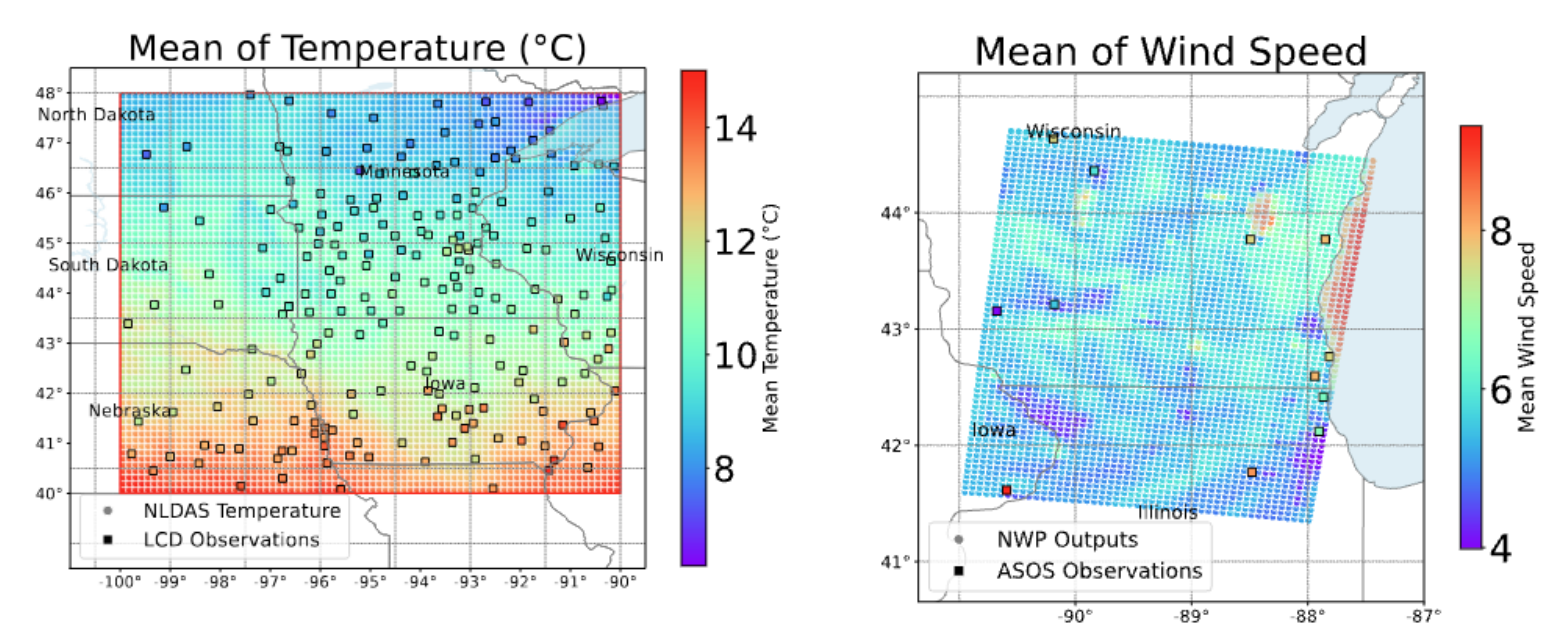}
    \caption{\textbf{Left: }The highlighted region in the Upper Midwest is enclosed within the bounding box. NOAA LCD stations are represented by squares, and the mean temperatures are displayed. \textbf{Right: }The region of interest surrounding Lake Michigan is depicted, with ASOS stations represented by squares. The temporal mean wind speed is displayed for all NWP simulation locations as well as ASOS observation sites.}
    \label{fig:subregion}
\end{figure}

Reanalysis temperature data are extracted from the North American Land Data Assimilation System (NLDAS-2) \cite{mitchell2004multi,Xia.2012} and are freely available at \url{https://ldas.gsfc.nasa.gov/nldas/v2/models}. NLDAS data is quality-controlled and available at a grid spacing of 12 km and an hourly temporal resolution. 
The study focuses on the Upper Midwest region of the United States, using data from the year 2010. We select an area of $64 \times 80 = 5120$ gridpoints, see left panel of Figure \ref{fig:subregion}. 

We choose the local climatological data (LCD) from the National Centers for Environmental Information (NCEI) of the National Oceanic and Atmospheric Administration (NOAA), which are available across the United States and its territories. Data can be freely downloaded  at \url{https://www.ncdc.noaa.gov/cdo-web/datatools/lcd}. The selected stations have sub-hourly measurement records at different times of the day throughout the year and across stations. Consequently, we make the measurement format uniform by taking the mean value over one-hour intervals for each station, ensuring consistency across the dataset.
Left panel of Figure~\ref{fig:subregion} displays the gridded NLDAS data for the region of interest, with mean temperature values represented on the map. The 213 LCD observation stations are marked with square symbols.

\subsection{Surface Wind Speed Data}
We consider the surface wind speed data studied in \cite{bessac+ca18}, which examines the  Lake Michigan region, covering parts of Wisconsin, Illinois, Indiana and Michigan. The presence of Lake Michigan significantly influences wind conditions and their modeling by numerical models. For consistency of statistical properties, the dataset is divided into three spatial clusters of 11, 12, and 8 stations, respectively. In this work, we focus the analysis on Cluster 2 in the region shown in Figure 1 of \cite{bessac+ca18}, see right panel of Figure \ref{fig:subregion}.

The observational data are sourced from the publicly available Automated Surface Observing System (ASOS) network (\url{ftp://ftp.ncdc.noaa.gov/pub/data/asos-onemin}) and consists of 1-minute wind speed data, discretized in integer knots (where one knot is approximately 0.5 m/s). Data are filtered with a moving window to smooth the discretization effects and picked every hour. 

The numerical model outputs are produced using weather research forecast (WRF) v3.6 \citep{skamarock+kdgbdwp08}, a state-of-the-art NWP system designed for both operational forecasting and atmospheric research. WRF provides a detailed representation of atmospheric physics, including cloud parameterization, land-surface models, atmosphere-ocean coupling, and radiation models. 
The NWP simulations are initialized daily during January 2012 using the North American Regional Reanalysis fields dataset, covering the Great Lakes region on a 5$\times$5 km grid with a 10-minute time resolution. The mean wind speed of the 12 observation locations and of the NWP outputs is shown in the right panel of Figure \ref{fig:subregion}. 

\section{Correction of Surface Air Temperature}\label{sec:temperature_results}

\subsection{Analysis of Path Signature Outputs}\label{sec:sig_analysis}
In this section, we analyze the path signature outputs when Algorithm~\ref{alg:SigPCA} is applied to the NLDAS gridded temperature data. 
The dyadic window depth is set to $d_w=3,$ which determines the level of recursive partitioning applied to the data. This hierarchical subdivision allows us to capture localized temporal variations. Additionally, we are considering only first-order integrals when computing the path signatures, i.e. depth-1 signatures. Unlike the window depth, which controls how the data is segmented, the signature depth dictates the level of interaction between variables in the extracted features. 
It is important to emphasize that the Sig-PCA is applied exclusively to the NLDAS model outputs at the gridpoints, not to the observational LCD data. With 5120 gridpoints, computing the path signatures over 7 sub-windows, as shown in Figure~\ref{fig:pathsignatures_subregion}, results in a total of 35,840 features (5120 $\times$ 7). PCA is then used to reduce the dimensionality of this dataset to 135 principal components explaining $99.5\%$ of the variation in the path signatures. These components capture both the temporal and spatial variability within the NLDAS data, providing a more compact representation of the underlying patterns. In practice, only this reduced dataset needs to be moved and used as inputs to train reconstruction and correction NNs. 

\begin{figure}
    \centering
\includegraphics[width=\textwidth]{   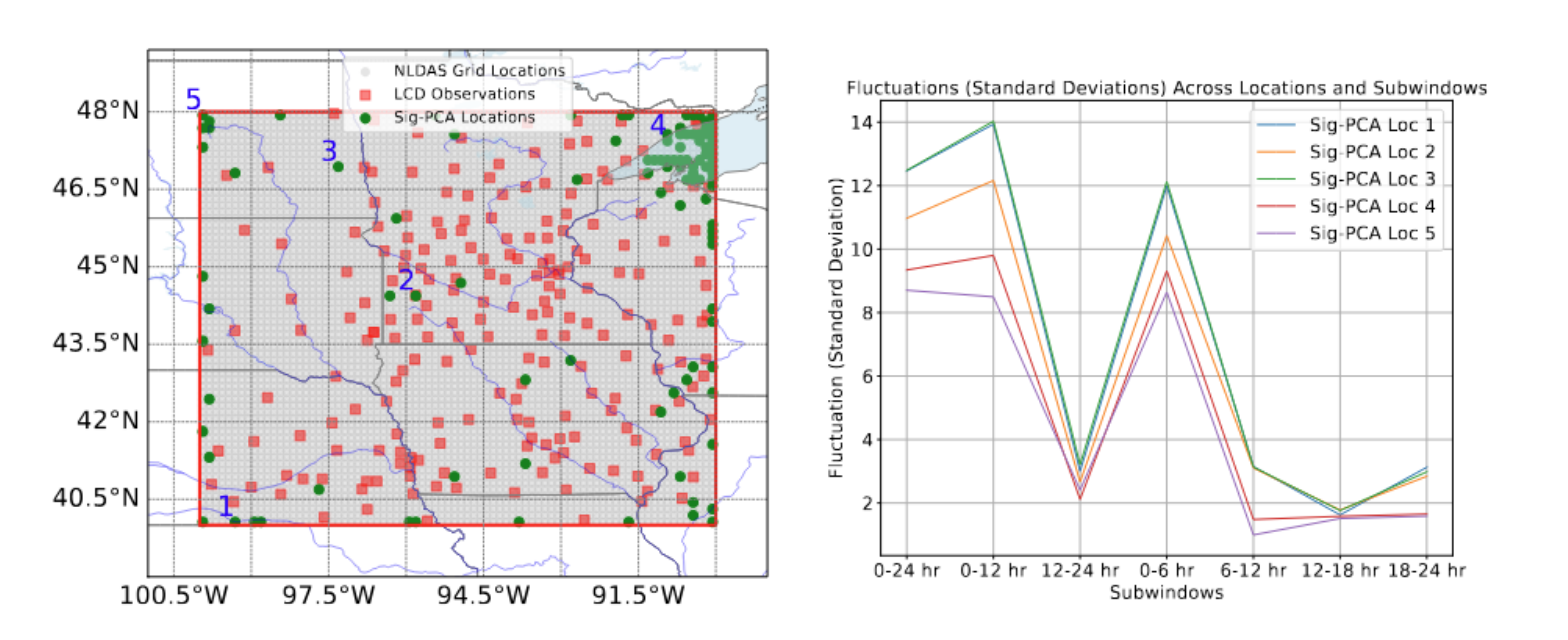}
    \caption{\textbf{Left: }Most contributing Sig-PCA locations are shown in green circles. Selected 5 stations are highlighted with blue numbers for further analysis. Blue curves represent rivers. Red squares represent LCD observation locations. \textbf{Right: }Fluctuations of path signatures of the selected 5 top Sig-PCA locations across various sub-windows. }
    \label{fig:topsigpca}
\end{figure}

\begin{figure}
    \centering
\includegraphics[width=0.85\textwidth]{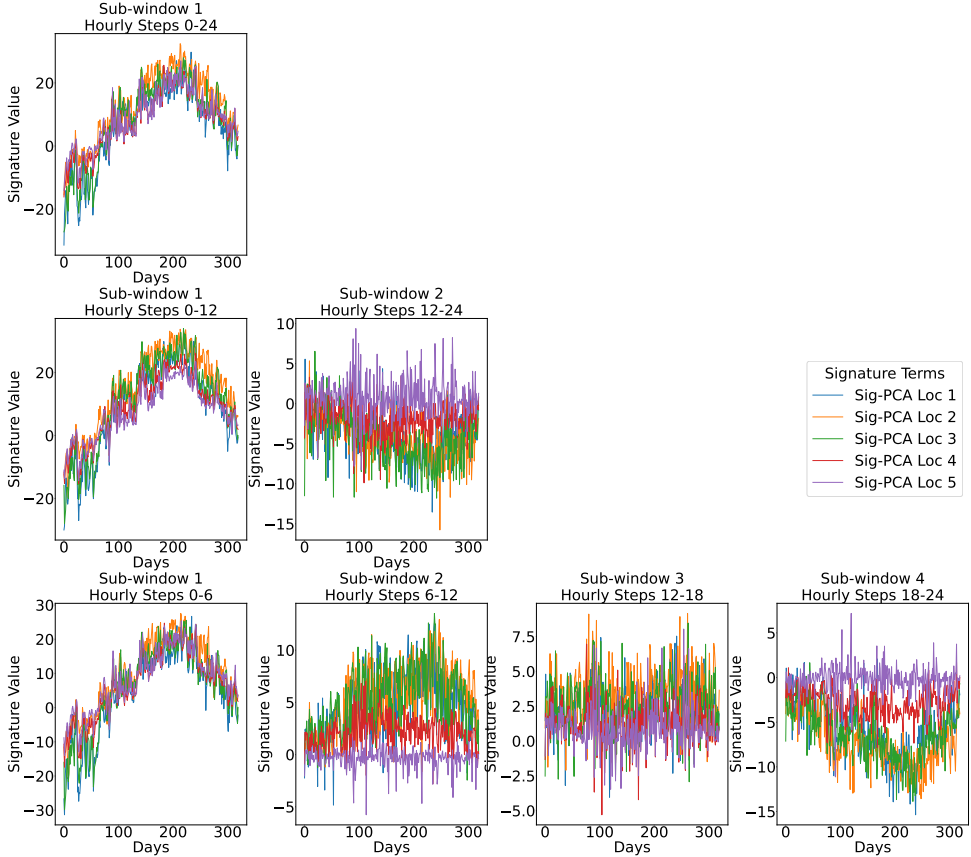}
    \caption{Path signatures variation across different sub-windows shown at the 5 selected locations (different colors).}
    \label{fig:pathsignatures_subregion}
\end{figure}

From these principal components, the top contributing locations are identified and highlighted as green circles in the left panel of Figure~\ref{fig:topsigpca}. The majority of these locations are concentrated over the lake in the upper right and along the boundaries of the region of interest. Although the temperature variance over the lake is low, as can be seen in the bottom panel A of Figure~\ref{fig:recon_mean_sd}, the locations over the lake still contribute significantly to the PCA. 
Additionally, we suspect that locations at the boundaries of the region of interest may be influenced by data truncation effects, which can introduce variability that is captured by PCA. This behavior is not observed in the wind data in Section 5~\ref{sec:wind_sig_analysis}, where most contributing locations are more centrally distributed.  We believe this is because temperature is a smoother field, making boundary regions relatively more informative than the interior. Among these, five key stations are selected  to highlight different data configurations and geographies.  Location 1 is located near a river. Location 2 is very close to an observation station. Location 3 is isolated from both the Sig-PCA and the observation locations. Location 4 lies along the lake-land boundary. Finally, Location 5 is the most remote among the selected points, being farthest from all observation stations. 

Figure~\ref{fig:pathsignatures_subregion} illustrates the differences in path signatures across various sub-windows at the highlighted locations, showcasing how the extracted features vary spatiotemporally providing insight into the underlying structure of the dataset. 
Each subplot corresponds to a different time segmentation of daily time windows, with the y-axis showing the signature values and the x-axis representing days across the year.  Signatures display higher values mid-year/late summer suggesting that the variability is more prominent during that period. This highlights the ability of path signatures to capture non-stationarity in time. Some locations exhibit more fluctuations than others in smaller sub-windows, reflecting spatial variability in the data, which is captured by path signatures. This figure shows how path signatures capture both seasonal and spatial variability. 

The right panel of Figure~\ref{fig:topsigpca} illustrates the variability in path signatures across the five locations, quantified as the standard deviation of the signatures, revealing distinct variations at different temporal scales as in Figure~\ref{fig:pathsignatures_subregion}. 
Locations 4 and 5 consistently show the least fluctuations across all sub-windows, indicating a more stable behavior throughout the day. In contrast, Locations 1 and 3 exhibit the highest fluctuations consistently across all time sub-windows, suggesting greater modeled variability in temperature at these locations. 
Fluctuations generally decrease as we move from the global window (covering all 24 hours) and the 0-12 hour sub-window to the later part of the day, i.e the 12-24 hour period. This suggests that temperature variability tends to reduce as the day progresses. However, fluctuations peak in the 0-6 hour sub-window, possibly reflecting changes in atmospheric conditions during the early morning hours. After this peak, fluctuations dip once more, but for Locations 1, 2, and 3, the fluctuations increase slightly during the 18-24 hour sub-window, suggesting a late-day increase in variability, possibly due to factors like evening cooling or shifts in wind patterns. 

\subsection{Reconstruction of NLDAS Model Outputs}\label{sec:res_recons}

The reconstruction NN is a fully connected feedforward regression model, consisting of 7 layers. The input layer is followed by successive hidden layers with progressively smaller sizes— 512, 256, 128, 64, 32, and 16 neurons—each followed by batch normalization and ReLU activation. The final output layer maps to the desired output size. The network's hierarchical structure allows it to capture complex relationships in the temperature dataset while batch normalization ensures stable training and improved convergence. The model is trained using mean squared error (MSE) loss for 500 epochs to ensure convergence and optimal performance optimizer learning rate $0.01$.

\begin{figure}
    \centering
\includegraphics[width=0.5\textwidth]{   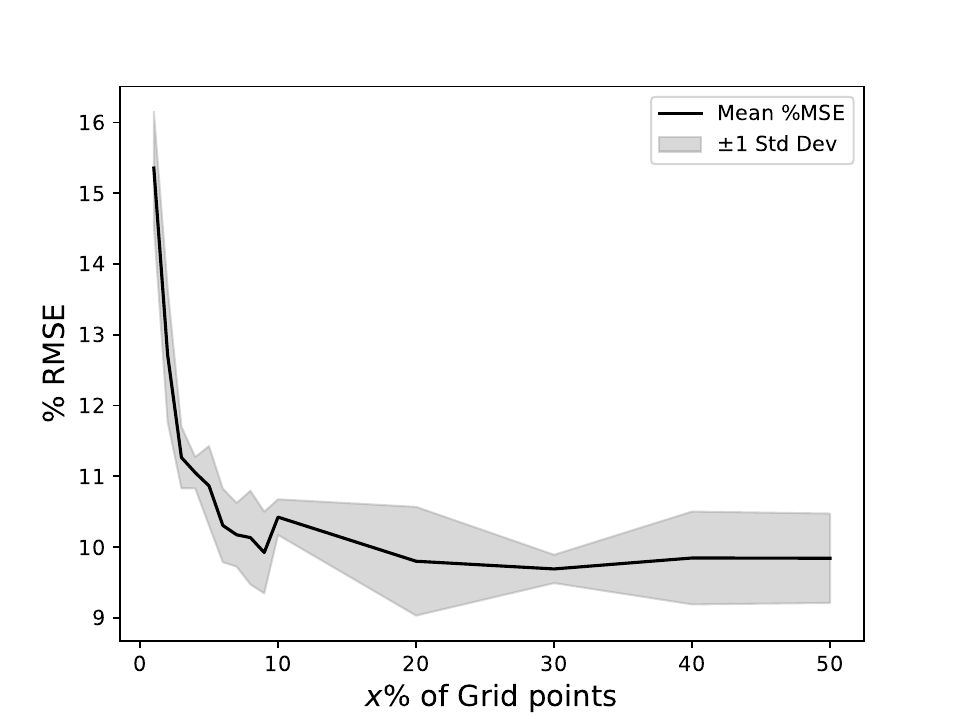}
    \caption{Sensitivity analysis of the percentage RMSE as a function of the percentage $x$ of gridpoints used for training the NN.}
\label{fig:sensitivity}
\end{figure}

As an initial validation step, we aim to only reconstruct  the NLDAS gridded data from the summary statistics from Section \ref{sec:sig_analysis} and without observation-informed corrections. 
We perform a sensitivity analysis to determine the number of training  gridpoints ($x$ from Figure \ref{fig:nn}) required to reconstruct the original NLDAS data from the summary statistics obtained using the Sig-PCA framework. 
After reconstructing the entire grid, the error between the original and reconstructed NLDAS is evaluated across all gridpoints. Figure~\ref{fig:sensitivity} presents the mean percentage RMSE, computed as $\% RMSE=\frac{RMSE(Predicted\  NLDAS, NLDAS)}{Mean(NLDAS)}\times 100$ as a function of the percentage $x$ of gridpoints used for training the reconstruction NN. 
To quantify the associated uncertainty, we conduct five independent simulations for each percentage of training points and compute the standard deviation, represented by the shaded gray area. The results indicate that the error stabilizes beyond 20$\%$, corresponding to approximately 1,024 gridpoints in the present case; however, we note that the $\%RMSE$  remains tolerable for smaller percentage of gridpoints as training outputs (below $16\%$ of $\%RMSE$ for $1\%$ of training output). 
Moreover, with a large grid, selecting $20\%$ of the gridpoints may result in a substantial amount of data. 
%

\begin{figure}
\centering
 \includegraphics[width=.9\textwidth]{   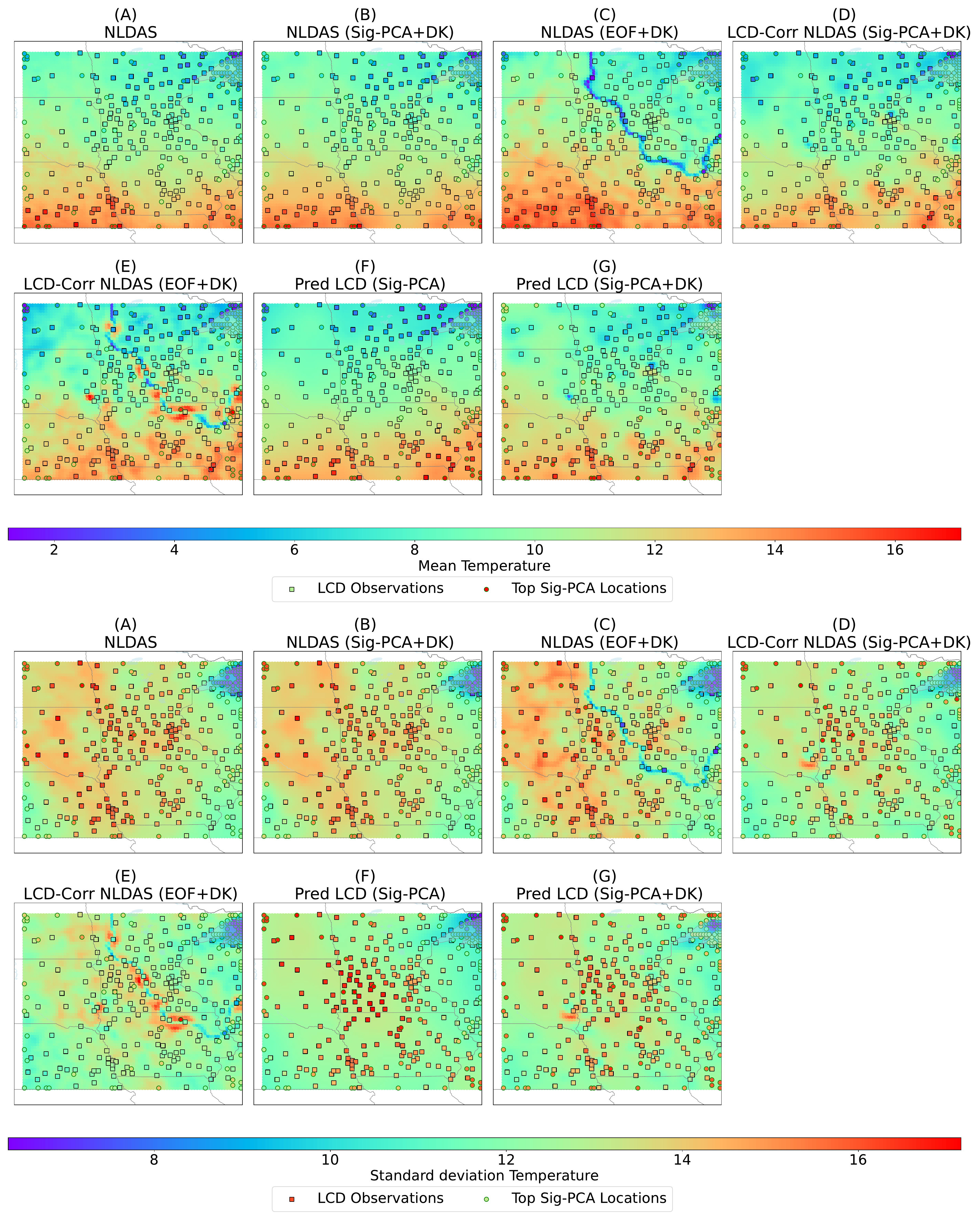}
  \caption{Comparison of different approaches capturing the mean ({\bf top}) and standard deviation ({\bf bottom}) of the original NLDAS (Panel A); reconstructed NLDAS without LCD correction using Sig-PCA and deep kriging (Panel B), and EOF and deep kriging (Panel C); LCD-corrected reconstructed NLDAS using Sig-PCA and deep kriging (Panel D), and EOF and deep kriging (Panel E); and outputted LCD using only Sig-PCA (Panel F), and Sig-PCA and deep kriging (Panel G).  Squares represent the LCD observation stations whereas the circles show the most contributing Sig-PCA locations.}
\label{fig:recon_mean_sd}
\end{figure}

Finally to mimic practical cases where limited amount of observational data are available as training outputs of the corrective NN, we evaluate the NLDAS reconstruction with $4\%$ of the total amount of gridpoints, yielding a corresponding RMSE of $11.05\%.$  The percentage $4\%$ arises from the number of  LCD stations representing around $4\%$ of the total number of NLDAS gridpoints.    
By training on  $4\%$ of the gridpoints (approximately 204 data points), we achieve a reasonable approximation of NLDAS data across the entire grid, comparing top (mean of reconstructions) and bottom (standard deviation of reconstructions) panels A and B in Figure \ref{fig:recon_mean_sd}. 
The Sig-PCA method (Panel B) outputs the model data well, although it misses some finer variations, likely due to relying only on first-order summary statistics. 

Several model comparisons are further conducted in Figure~\ref{fig:recon_mean_sd} and discussed in the following section:  
\begin{itemize}
 \item[1.] \textbf{NLDAS (Sig-PCA+DK)-} Reconstructive NN (Panel B) trained to reconstruct NLDAS with
 Sig-PCA summary statistics, spatial coordinates and deep-kriging basis functions as inputs.  
    \item[2.] \textbf{NLDAS (EOF+DK)-} Reconstructive NN with EOF modes as summary statistics (Panel C) as discussed in Section \ref{sec:eof}, with deep-kriging basis function and spatial coordinates as inputs.
    \item[3.] \textbf{LCD-Corr NLDAS (Sig-PCA+DK)-} Reconstructive + Corrective NN (Panel D) to
     predict LCD-corrected reconstructed NLDAS from Sig-PCA summary statistics, spatial coordinates, and deep-kriging basis functions inputs.  
    \item[4.] \textbf{LCD-Corr NLDAS (EOF+DK)-} Reconstructive + Corrective NN with EOF modes (Panel E). Similar to the previous method, but with EOF modes as summary statistics, spatial coordinates and deep-kriging basis functions as inputs.
    \item[5.] \textbf{Pred LCD (Sig-PCA)-} Direct Learning of LCD Observations (Panel F) from Sig-PCA summary statistics and  spatial coordinates as inputs, and LCD observations  as outputs.  
    \item[6.] \textbf{Pred LCD (Sig-PCA+DK)- } Direct Learning of LCD Observations with deep-kriging basis functions (Panel G) as additional inputs to Sig-PCA summary statistics and spatial coordinates.
\end{itemize}

\subsection{Evaluation of the Observation-Corrected Fields}

We first focus on evaluating the quality of outputting the LCD observations directly without intermediate NLDAS reconstruction (Panels F and G Figure \ref{fig:recon_mean_sd}). The observation-prediction only (Panel F) results in fields matching only at the observation stations without correction effects outside observation stations. When deep kriging is incorporated (Panel G), a radial interpolation effect around observation stations can be observed and allows for a smoother interpolation within the neighborhood of LCD stations. 
We next evaluate the quality of the observation-corrected reconstructed model outputs (Panels D and E). 
The mean LCD-corrected reconstructed NLDAS using Sig-PCA with deep kriging, (Panel D), aligns most closely with observation stations and provides more detailed spatial interpolation in their vicinity. However, in regions without LCD data, the reconstruction and correction lack accurate interpolation, underscoring the importance of LCD data in refining the model dataset. 
Finally, we compare our approach with EOF-based summary statistics. The first 11 EOF components are kept as summary statistics as they capture $99.5\%$ of the variance in the data. 
The EOF-based reconstruction introduces an overly dominant spatiotemporal artifact appearing as a blue curve, (Panel C). This effect is slightly diminished when LCD observations are incorporated (Panel E). The Sig-PCA approach outperforms the EOF-based approach by preserving spatial accuracy and reducing interpolation errors, resulting in a more reliable reconstruction and correction. 

\begin{figure}
    \centering
\includegraphics[width=\textwidth]{   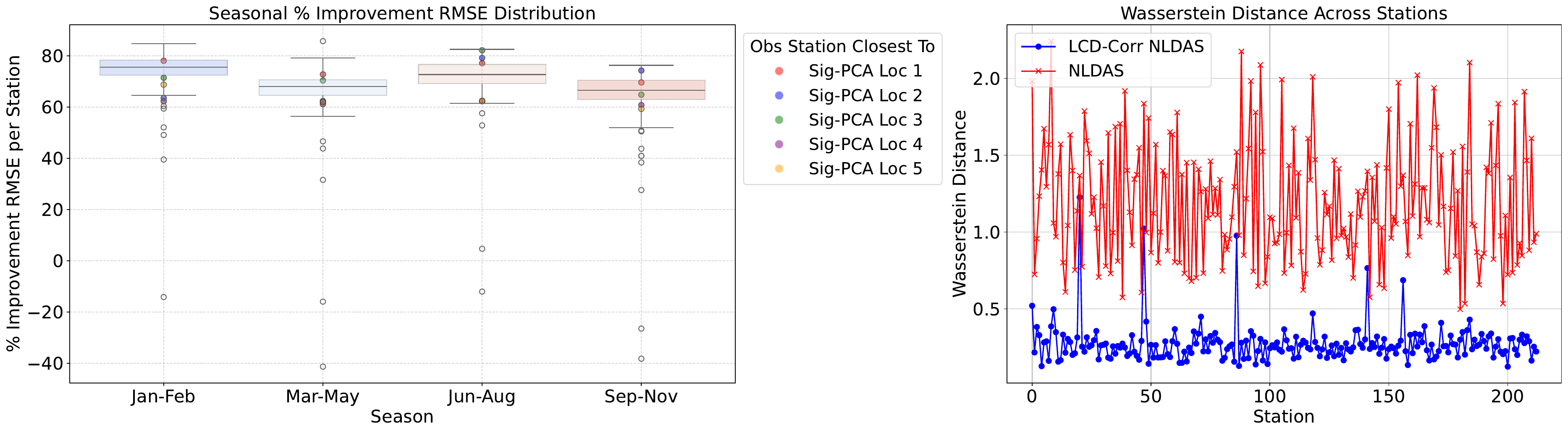}
    \caption{\textbf{Left: }Seasonal $\%$ Improvement RMSE boxplot of LCD-corrected NLDAS and NLDAS with respect to LCD observation stations. The stations closest to the Sig-PCA locations are highlighted in different colors. \textbf{Right:} 1-Wasserstein distance across stations the LCD observations and LCD-corrected NLDAS (blue lines) and NLDAS (red).}
    \label{fig:rmse}
\end{figure}

The left panel of Figure~\ref{fig:rmse} presents boxplots of the 
 $\%$ improvement RMSE computed as $\frac{RMSE(NLDAS,\  LCD)-RMSE(LCD\text{-}corr\  NLDAS,\  LCD)}{RMSE(NLDAS,\  LCD)}*100.$ This metric evaluates the improvement of the proposed LCD-corrected reconstructed NLDAS over the original NLDAS data. The higher the percentage, the more effective is our correction approach in aligning with the LCD observations compared to NLDAS, whereas a value less than 0 suggests that the original model outputs are closer to the observations than the proposed corrections. 
The observation stations that are closest to the five selected locations are highlighted, to evaluate the correction quality. Location 2 is kept for deeper analysis in the following as it is the closest to an observation station. Results for other Sig-PCA locations can be found in the supplemental material. 
As shown in the left panel of Figure~\ref{fig:rmse}, most values are above $60\%,$ indicating a substantial improvement in recovering LCD observations over original NLDAS outputs. A few observation stations exhibit values below zero; upon further investigation, all these stations contain significant amounts of missing data, hence providing poor corrections. 
The $1$-Wasserstein distance \citep{muskulus2011wasserstein} between the LCD observations and both the NLDAS (red) and the LCD-corrected NLDAS (blue) are shown on the right panel of Figure~\ref{fig:rmse}. 
The $p$-Wasserstein distance provides a distance between probability distributions and, in the one-dimensional case, as here, it corresponds to the $\mathcal{L}^{p}$-norm between the quantile function  of the two distributions at stake. 
The 1-Wasserstein distances of the LCD-corrected NLDAS consistently exhibit lower values than the NLDAS one, highlighting that the distributions of LCD-corrected NLDAS are closer to the LCD-observation distribution than the original NLDAS ones. 

\begin{figure}
    \centering
\includegraphics[width=\textwidth]{   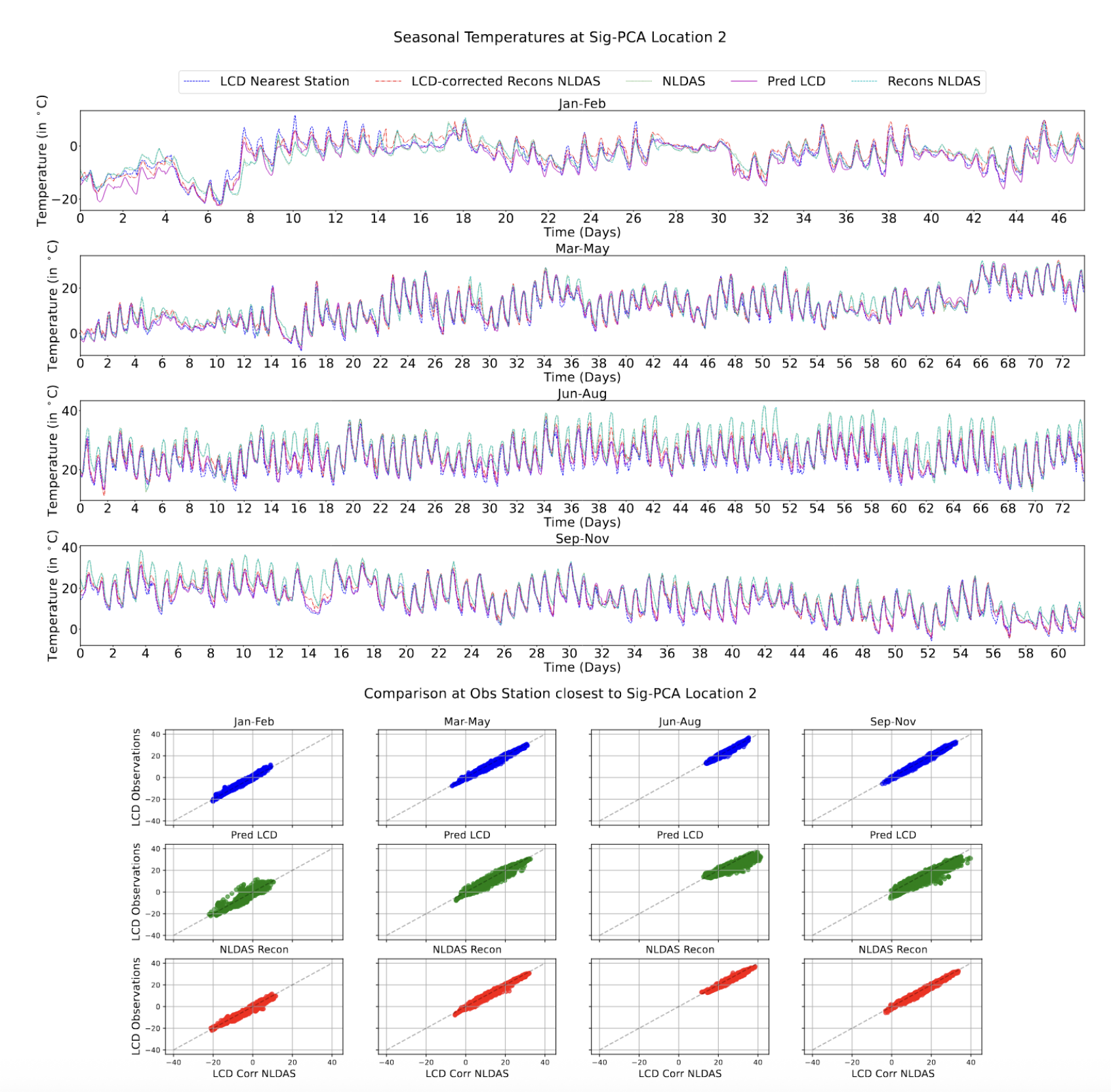}
    \caption{\textbf{Top:} Comparison of different time series  for Location 2 with the time series of nearest observation station, located very nearby. The LCD-corrected NLDAS is denoted in red, NLDAS in green, directly outputted LCD in purple and reconstructed NLDAS in cyan. Different rows represent different seasons. \textbf{Bottom:} QQ-plot of outputs for different approaches (x-axis) compared with the LCD observations (y-axis) at the nearest observation station to Sig-PCA at Location 2. The rows represent predicted LCD (top in blue), reconstructed NLDAS (middle in green), and LCD-corrected reconstructed NLDAS (bottom in red), while the columns correspond to the four different seasons.}
    \label{fig:6loc2}
\end{figure}

Top panels of Figure~\ref{fig:6loc2} present a comparative analysis of time series at Location 2 across different seasons. Time series include the original NLDAS, LCD observations from the nearest station, directly outputted LCD data at the target location, reconstructed NLDAS at the same location, and the final LCD-corrected reconstructed NLDAS. The reconstructed NLDAS (cyan) is largely similar to the original NLDAS (green). Additionally, both the directly outputted LCD (purple) and the LCD-corrected NLDAS (red) generally resemble the LCD (blue) at the nearest station across most seasons. However, capturing accurate winter patterns is challenging. In early January, the LCD corrections align well with observations, while the direct LCD output tends to underestimate the observed temperature. By mid-January, the directly outputted LCD matches the LCD more closely than the corrections, highlighting the complexity in winter temperature variations.

Bottom panels of Figure~\ref{fig:6loc2} show the quantile-quantile (QQ) plots of the outputs for different approaches (x-axis) compared with LCD observations (y-axis) at the observation stations. We observe that both predicted LCD (top row) and LCD-based corrections (bottom row) closely match the LCD observations, even in the tails that are typically more challenging to capture, highlighting the importance of LCD observations to correct NLDAS model outputs.

\begin{figure}
    \centering
\includegraphics[width=\textwidth]{   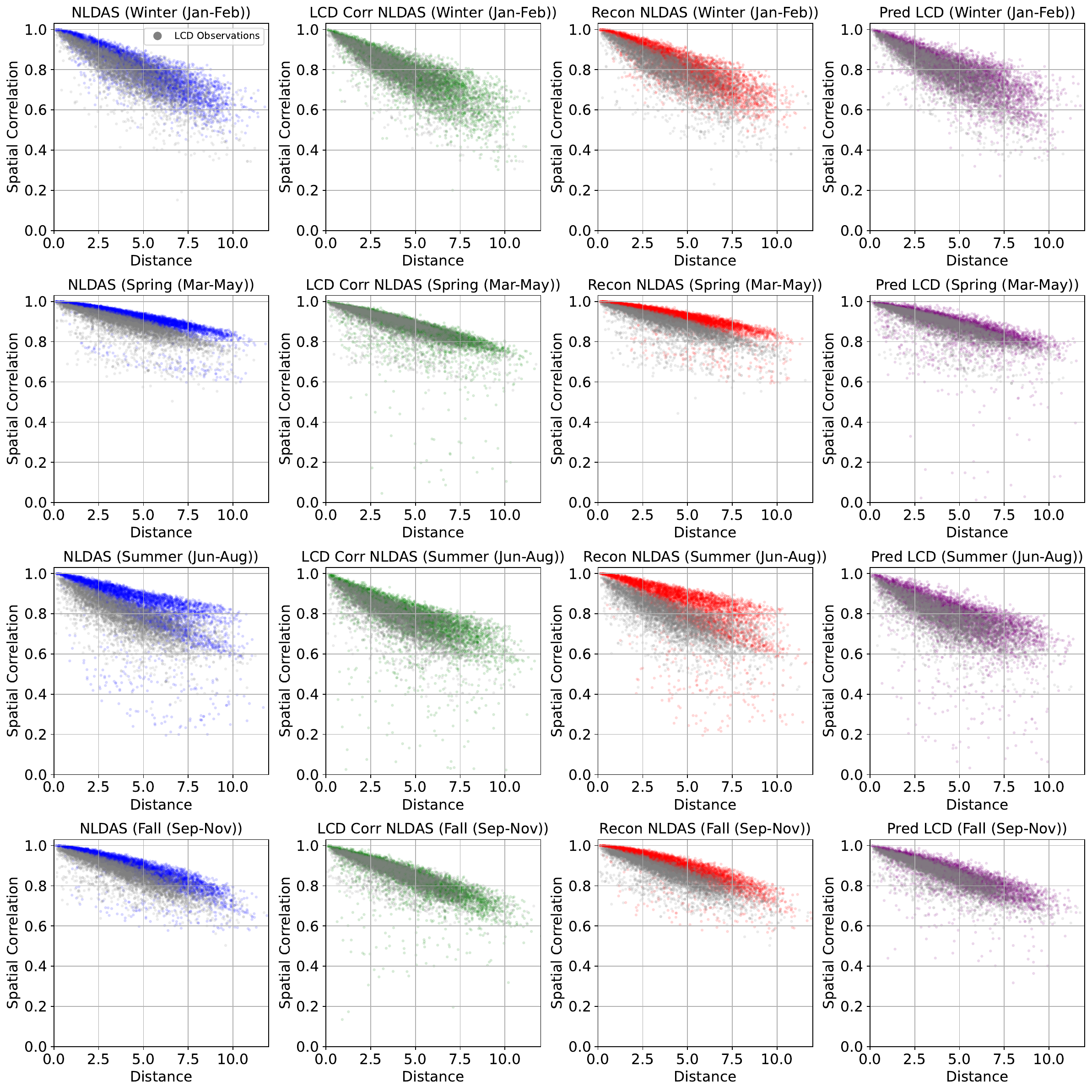}
    \caption{Correlation as a function of distance between gridpoints. {\bf Rows} depict different seasons and {\bf columns} different data from {\bf left} to {\bf right}: original NLDAS data, LCD-corrected reconstructed NLDAS, NLDAS reconstruction, and directly predicted LCD. Correlation between LCD stations is shown in grey. }
\label{fig:nldascorrelation}
\end{figure}

Figure~\ref{fig:nldascorrelation} shows the spatial correlation as a function of the distance between gridpoints for each season. In winter, a weaker spatial dependence is observed, while spring exhibits a stronger and more spatially coherent correlations. Fall follows an intermediate pattern, with correlation higher than in winter but lower than in spring. Summer has a more spread-out correlation that lies between fall and winter. 
The reconstructed NLDAS (red) retains the original NLDAS seasonal correlation structure, reflecting the model output's inherent spatial dependencies. However, when LCD observations are incorporated—either through direct prediction of LCD values (purple) or via the correction-based approach (green)—the correlation patterns of both methods match closer the LCD-observation correlations (grey). 
Finally, the correction-based approach (green) shows correlation closer to the LCD-observation ones compared to the directly predicted LCD (purple), highlighting the need to integrate both model outputs and observations for most accurate representations of spatial patterns. 

\begin{figure}
    \centering
\includegraphics[width=0.8\textwidth]{   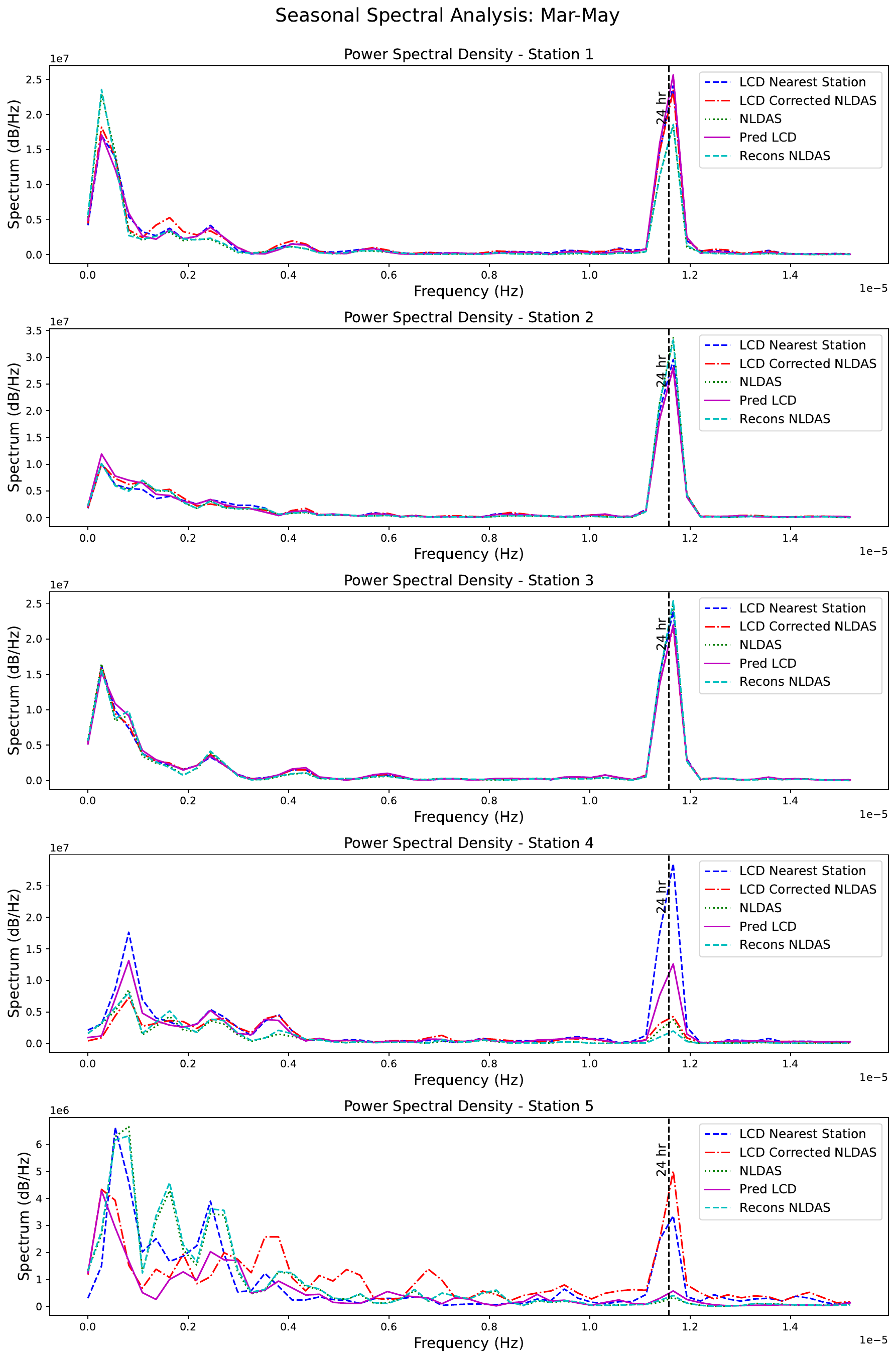}
    \caption{Seasonal spectrum for spring (March-May) for the 5 selected locations across the rows. The blue dotted lines represent the LCD at the nearest observation station, red line corresponds to the LCD-corrected NLDAS, the green line represents the original NLDAS, the purple line denotes the directly outputted LCD, and the cyan line indicates the reconstructed NLDAS. }
    \label{fig:marchmayspectra}
\end{figure}

Figure \ref{fig:marchmayspectra} presents the seasonal power spectral densities for spring at the five selected locations, providing insights into the frequency characteristics of the different approaches. 
Location 1 exhibits both a dominant 24-hour peak and a lower-frequency peak, indicating a combination of diurnal and slower variations. Locations 2, 3, and 4 show a strong 24-hour peak with minimal lower-frequency influence, suggesting  daily cycles to be primary drivers of variability. In contrast, Location 5 demonstrates multiple dominant low-frequency peaks, implying that multiple slower patterns play a significant role in variations at this location. At Location 1, the spectra of the LCD-corrected NLDAS (red) and directly outputted LCD (purple) closely resemble the LCD-observation one (blue). However, around frequency of 1.5 days, the LCD-corrected NLDAS exhibits a slightly more prominent peak than other methods, suggesting that the correction process amplifies certain longer-term variations present in NLDAS (green), which are not as strongly reflected in the directly outputted LCD. Locations 2 and 3 exhibit a high level of matching between the five compared outputs. At Location 4, the directly outputted LCD replicates the nearest LCD observation station in both the daily cycle and lower-frequency patterns. The LCD correction method exhibits a lower peak, reflecting a weaker daily cycle intensity and low-frequency behavior compared to the direct LCD output. This is mainly due to the NLDAS dataset itself having a lower peak at these frequencies, so when LCD correction is applied, it incorporates NLDAS characteristics, leading to dampened diurnal and long-term variability. Finally, Location 5 is the one farthest from LCD-measurement stations, justifying the increased discrepancy between measurements and model outputs, and highlighting the spatial diversity in the frequency content. This location exemplifies the limitations of observation-based correction when these latter are sparse despite using kriging.

In conclusion, the proposed Sig-PCA approach effectively extracts lower-dimensional path signature summaries to reconstruct model outputs. While leveraging these summaries, localized observation corrections are also learnt; integrating both data sources to produce more reliable and accurate estimates. In particular, most statistics used in validation
are matching the observational ones when the Sig-PCA is used in reconstruction and correction setting, highlighting the need to embed information from both model and measurements.

\section{Correction of Surface Wind Speed}\label{sec:wind_results}

\subsection{Analysis of Path Signature Outputs}\label{sec:wind_sig_analysis}
The NWP dataset consists of surface wind speed structured as a space-time array of size $(49, 68, 695),$ with $49 \times 68 = 3,332$ spatial gridpoints, and $695$ time-steps. To increase the sample size for NN training, simulations  are reorganized into 15-hour intervals (46 samples) rather than daily 24-hour segments (29 samples). 
The depth-1 path signature is computed at each of the 3,332 gridpoints of the NWP model outputs for a dyadic window of depth 3. As previously, PCA is applied on the resulting path signature, retaining 41 components that explain 99.5$\%$ of the total variance. 
\begin{figure}
    \centering
   \includegraphics[width=0.7\textwidth]{   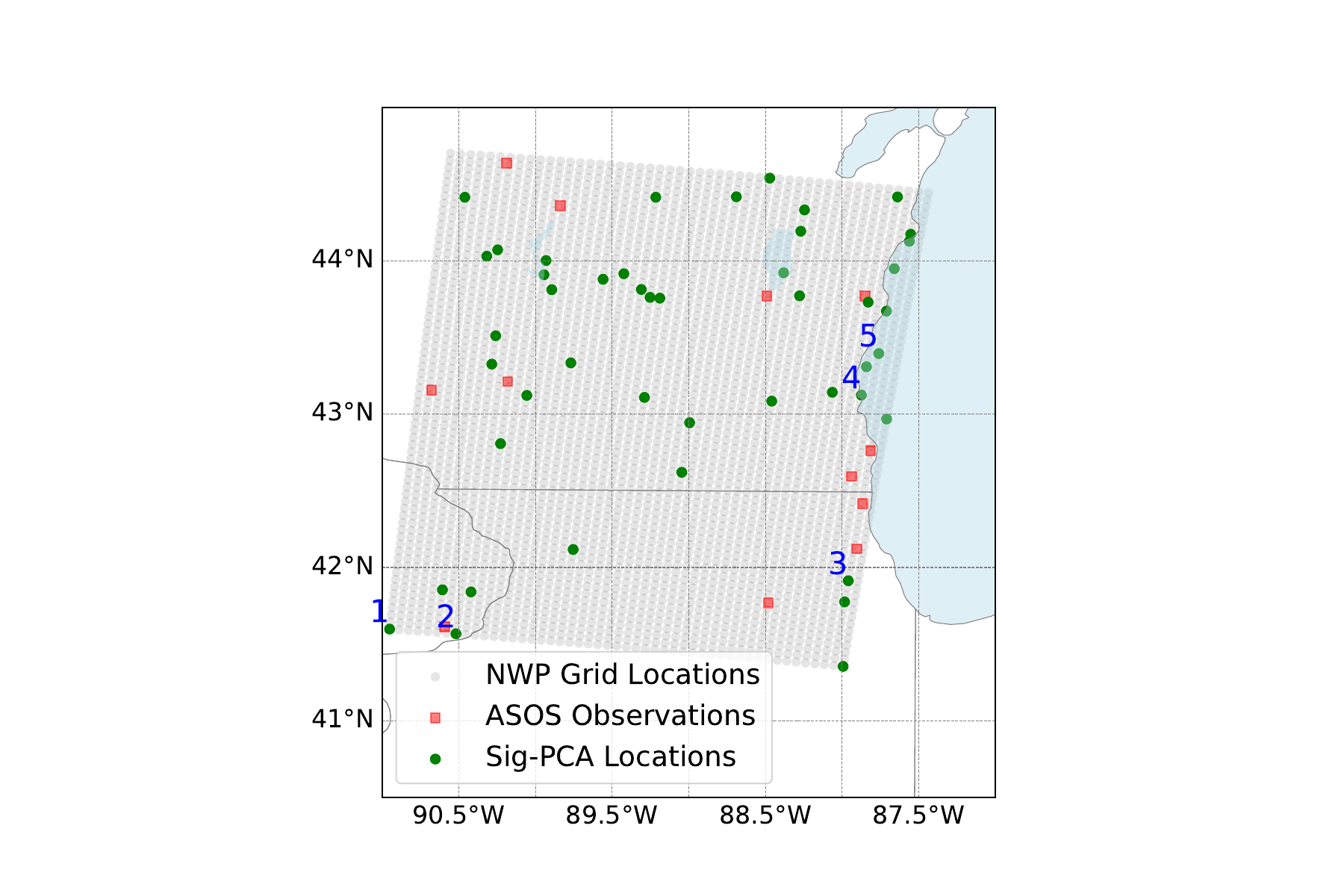}
    \caption{Grey dots mark the original NWP grid, the green circles show the top contributing locations to the Sig-PCA step, and the red squares show the ASOS observation stations.}
    \label{fig:nwplocationsmap}
\end{figure}
Many of the top-contributing  Sig-PCA locations are concentrated around water bodies, see Figure~\ref{fig:nwplocationsmap}, where both the modeled mean wind speed and variance are higher, see Panel A in Figure~\ref{fig:nwp_meancomparison}. 
From these, we select five specific locations, each representing distinct model characteristics. Location 1 is positioned at the farthest corner of the grid and reasonably close to an ASOS measurement station. Location 2 is very close to an ASOS observation station, allowing for direct comparison with observational data. Location 3 is near an urban center, Chicago. Location 4 lies along the water-land boundary. Finally, Location 5 is situated on Lake Michigan, representing water-body wind dynamics. 

\begin{figure}
    \centering
    \includegraphics[width=\textwidth]{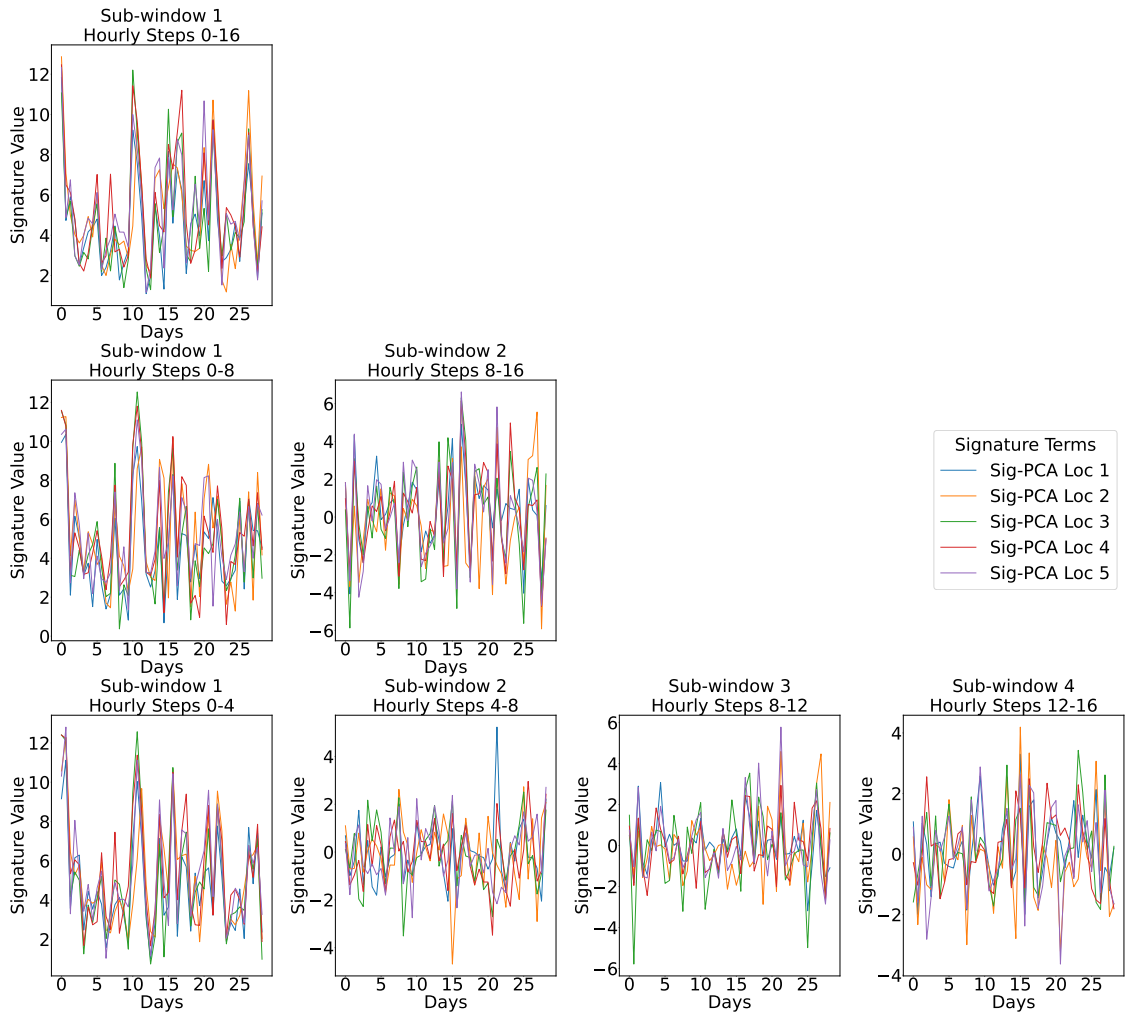}
    \caption{NWP depth-2 path signatures of the five selected locations across various sub-windows. }
    \label{fig:nwpsighier}
\end{figure}

\begin{figure}
    \centering
    \includegraphics[width=\textwidth]{   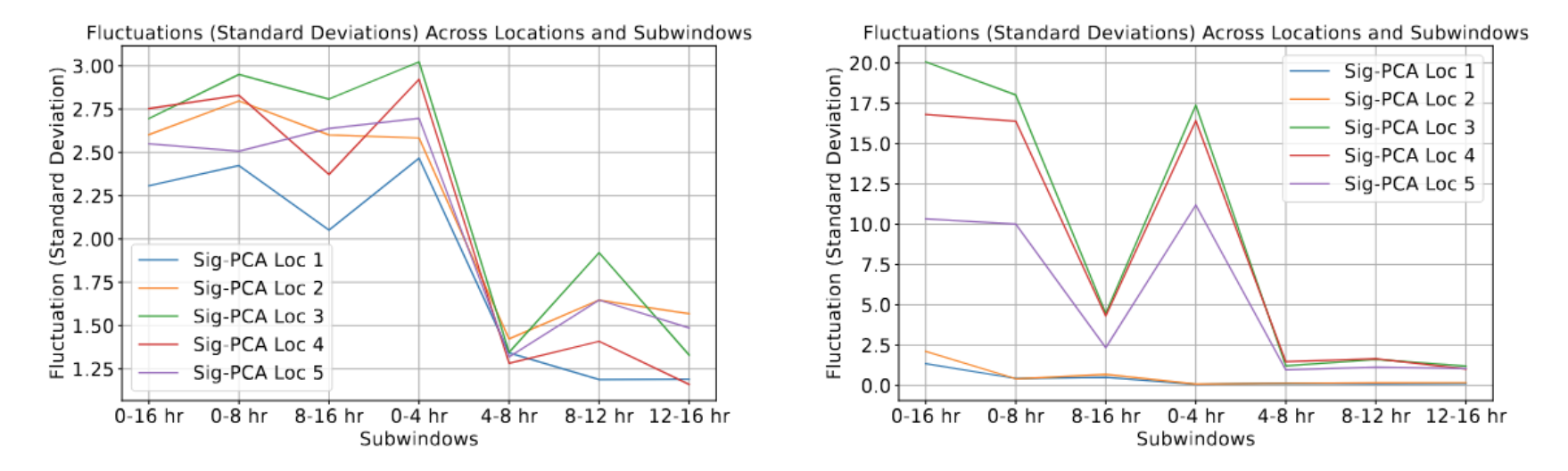}
    \caption{Fluctuations of path signatures of the NWP wind data at the five selected locations across various sub-windows. {\bf Left:}  depth-1 path signatures, {\bf right:} depth-2 path signatures on the five most contributing stations of depth-1 Sig-PCA. }
    \label{fig:nwpsd}
\end{figure}

Figure~\ref{fig:nwpsighier} displays the path signatures at the five selected locations, showing their variations across different sub-windows. 
In Figure~\ref{fig:nwpsighier}, since wind speed is studied over a month, it does not contain seasonal and longer-term trends that could also be seen in the corresponding path signatures. In comparison, Figure \ref{fig:pathsignatures_subregion}, studying temperature across a year, reveals seasonal patterns captured by the path signatures. As opposed to Figure \ref{fig:pathsignatures_subregion}, the displayed locations exhibit roughly similar trends in their depth-1 path signature. 
The left panel of Figure~\ref{fig:nwpsd} presents a plot of the standard deviation across the different sub-windows for all five locations, highlighting the spatial and temporal variability in the modeled wind dynamics. Location 1 consistently exhibits the least fluctuation across all sub-windows, suggesting relatively stable modeled wind patterns. In contrast, Location 3 shows the highest fluctuations, indicating more volatile wind conditions. 
Locations 1-4 experience a dip in fluctuations between the first two eight-hour periods, whereas Location 5 shows an increase, likely due to its unique positioning over Lake Michigan, where water-body wind dynamics are modeled differently by the numerical model compared with land-based locations. Between the 8 and 12 hour period, all locations except Location 1 exhibit a peak in wind speed fluctuations, while Location 1 shows a dip, reinforcing its relatively stable wind characteristics. 

\begin{figure}
    \centering
\includegraphics[width=\textwidth]{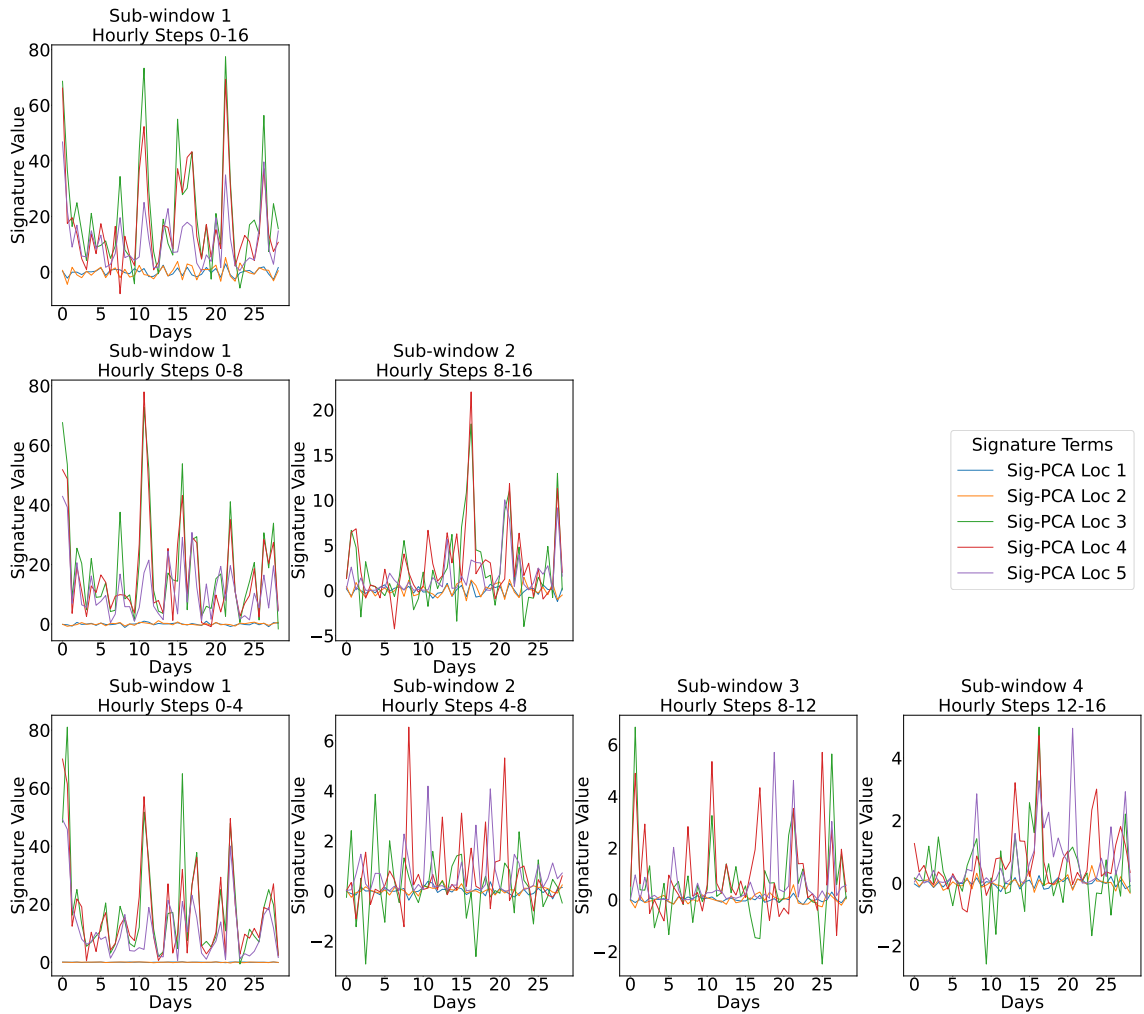}
\caption{NWP depth-2 path signatures of the selected locations across various sub-windows.}
    \label{fig:nwpdepth2}
\end{figure}

To explore higher-order content in the NWP wind dataset, we apply a depth-2 signature to the top Sig-PCA locations identified in the depth-1 step, avoiding applying depth-2 Sig-PCA uniformly across all locations.  Depth-2 fluctuations and path signatures are shown, respectively, in the right panel of Figure~\ref{fig:nwpsd} and Figure~\ref{fig:nwpdepth2}. 
Contrary to depth-1 signatures, depth-2 signatures exhibit different patterns across locations. 
The differences arise because depth-1 captures broader temporal trends, while depth-2 reflects higher-order variations within the wind dynamics. 
Rather than solely considering the magnitude of change in depth-1 signatures, depth-2 signatures encode how different time steps influence each other, reflecting more complex interactions. 
Locations 3 and 4 likely exhibit higher depth-2 fluctuations likely due to increased local variability in that area. 
In contrast, Locations 1, and 2 demonstrate more stable patterns, indicating less pronounced higher-order wind speed interactions. 

\subsection{Reconstruction of NWP Model Outputs}\label{sec:nwp_res_recons}
The NN architecture used for these datasets consists of two hidden layers, with 128 and 64 neurons respectively, followed by batch normalization and a ReLU activation, and finally an output layer. Unlike the previous NN architecture, ReLU is applied to the final output to ensure non-negative wind speed values. The model is trained using the MSE loss for 500 epochs and optimized with the Adam optimizer with an optimal learning rate of 0.01. 

We evaluate the effectiveness of different NN inputs in capturing spatial and temporal variability of wind speed in Figure~\ref{fig:nwp_meancomparison}, specifically comparing: 
\begin{itemize}
 \item[1.] \textbf{NWP (Sig-PCA1+DK)- }  (Panel B). 
     Reconstructing NWP with depth-1 Sig-PCA, spatial coordinates and deep kriging basis functions.
     
    \item[2.] \textbf{NWP (Sig-PCA2+DK)- }  (Panel C). 
    Similar to above, but with depth-1 and depth-2 Sig-PCA at selected locations.
    
    \item[3.] \textbf{NWP (EOF+DK)- }  (Panel D). 
Reconstructing NWP with EOF as summary statistics, spatial coordinates and deep-kriging basis functions as input.

    \item[4.] \textbf{ASOS-Corr NWP (Sig-PCA1+DK)- }  (Panel E). 
    Reconstructive + Corrective NN with depth-1 Sig-PCA, spatial coordinates and deep-kriging basis functions.
     
    \item[5.] \textbf{ASOS-Corr NWP (Sig-PCA2+DK)- }  (Panel F). 
    Similar to above, but with both depth-1 and depth-2 at selected locations.
    
    \item[6.] \textbf{ASOS-Corr NWP (EOF+DK)- }  (Panel G).  
     Same to above, but EOF as summary statistics with spatial coordinates, deep-kriging basis functions as inputs. 
\end{itemize}

In the EOF-based approach, 17 modes explain $99.5\%$ of variation in the dataset. By leveraging these modes as summary statistics, the EOF method provides a different perspective on variability representation, focusing on dominant patterns from the NWP wind data. 

Top plots of  Figure~\ref{fig:nwp_meancomparison} present a comparison of the mean wind speed estimates across different methods, while the bottom plots illustrate their standard deviation. We first evaluate the quality of the reconstructions only (without observational correction). The underlying structure of the NWP (Panel A) is retained in the reconstructions from depth-1 and depth-2 Sig-PCA approach (Panels B and C respectively). However, the EOF-based reconstruction (Panel D) introduces localized areas of higher mean wind speed and standard deviation that are not present in the original NWP data.

\subsection{Evaluation of the Observation-Corrected Fields}

\begin{figure}
\centering
\includegraphics[width=0.85\textwidth]{   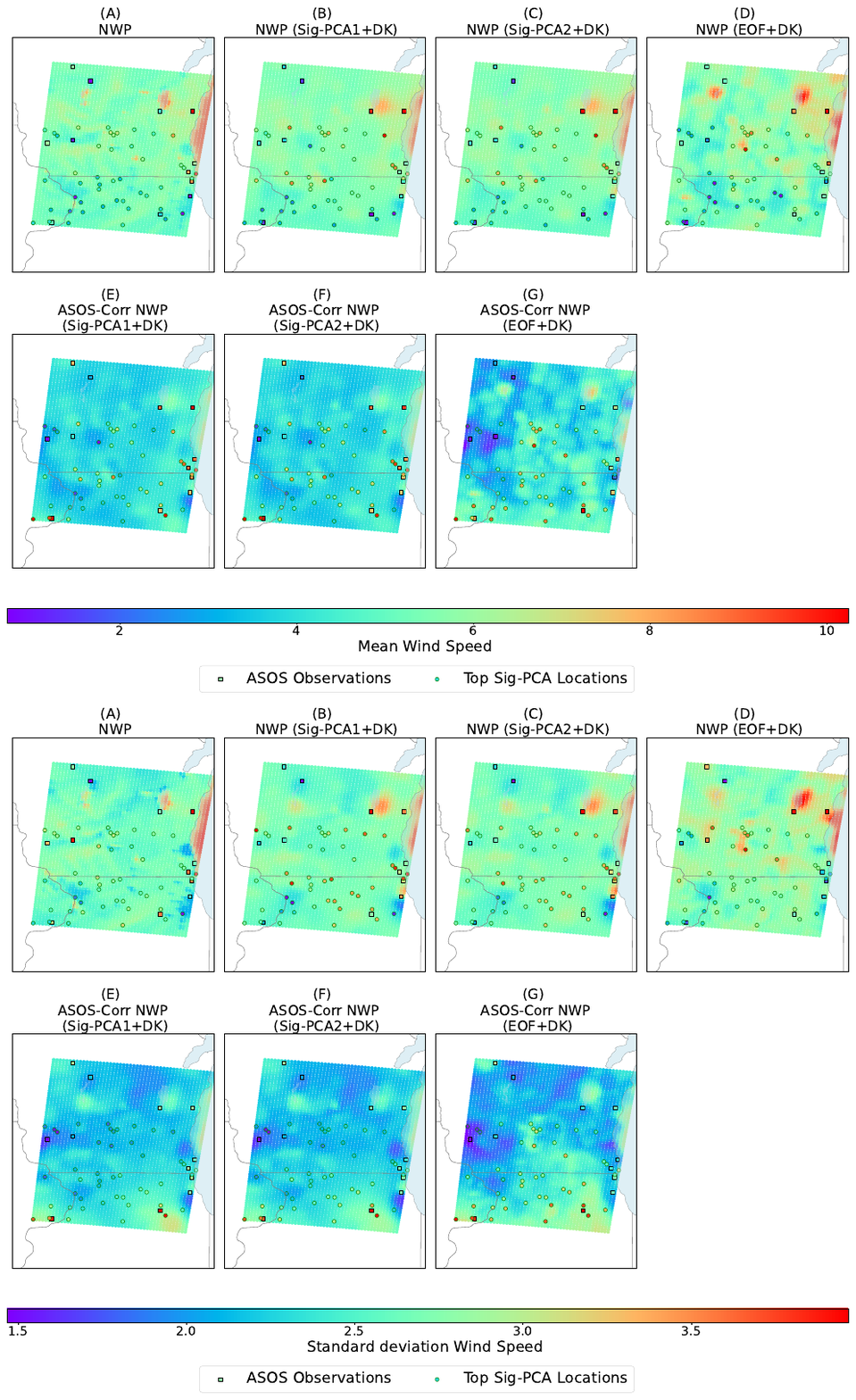}
\caption{Comparison of different approaches of the mean ({\bf top}) and standard deviation ({\bf bottom}) of the original NWP (Panel A); reconstructed NWP without ASOS correction using Sig-PCA1 with deep kriging (Panel B), Sig-PCA2 with deep kriging (Panel C), and EOF and deep kriging (Panel D); and ASOS-corrected reconstructed NWP using Sig-PCA1 and deep kriging (Panel E), Sig-PCA2 and deep kriging (Panel F), and EOF and deep kriging (Panel G). Squares represent the ASOS observation stations whereas the circles show the most contributing Sig-PCA locations. }
\label{fig:nwp_meancomparison}
\end{figure}

While the underlying structure for NWP reconstruction only remains consistent across both depth-1 and depth-2 (Panels B and C of Figure~\ref{fig:nwp_meancomparison}), the correction patterns are more pronounced for the depth-2 approach (Panels E and F) when observations are included.   
The differences are observed in the standard deviation plot in Panels E and F, where for correction, depth-1 Sig-PCA exhibit higher standard deviation in certain regions, indicating more estimated variability not exhibited by the station data. In contrast, depth-2 Sig-PCA corrections display a lower standard deviation in these same areas, showcasing the increased sensitivity of depth-2 in revealing subtle variations.
When incorporating ASOS observations for correction (Panels E and F), both the mean and standard deviation decrease significantly, aligning with the observed conditions that tend to be lower in these regions. This highlights the effectiveness of embedding observational data into the framework and higher-order depth in the signature in the case of wind data.  
When EOFs are used as summary statistics for correction (panel G), several clusters of elevated mean wind speeds are observed and interpreted as artifacts because they are absent from the model or observational data. The standard deviation exhibits less refinement and more dispersion than in the Sig-PCA approach. This likely occurs because EOF emphasizes  dominant modes that capture the largest variance, leading to the representation of broad features without  fine or localized features that Sig-PCA effectively captures.

\begin{figure}
    \centering
\includegraphics[width=\textwidth]{   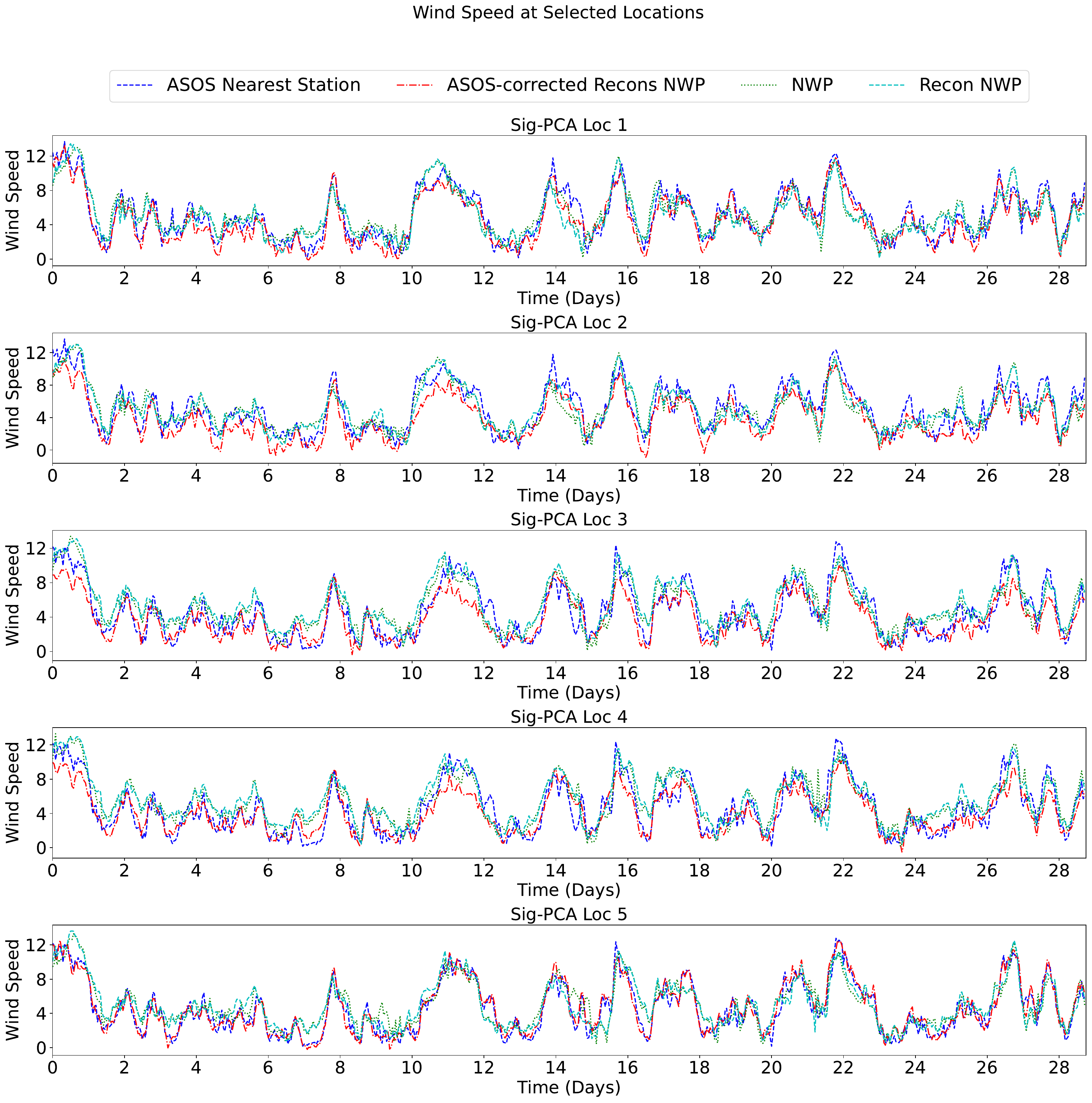}
    \caption{Comparison of different time series across all locations. The ASOS-corrected NWP is denoted in red, NWP in green, reconstructed NWP in cyan and ASOS at nearest station in blue.}
    \label{fig:nwptimeseries}
\end{figure}

Figure~\ref{fig:nwptimeseries} compares the gridded data's NWP time series (green),  reconstructed NWP (cyan), ASOS-corrected reconstructed NWP (red) and the nearest station's ASOS data (blue). For Sig-PCA locations 1-3, the NWP and reconstructed NWP are very similar, with corrections closely aligning with the nearest ASOS observation station.  However, at Location 4, the corrections do not fully match the nearest observation station between days 10-12, highlighting the approach's trade-off between model and observational data when the latter are farther away from locations of interest. Similarly, for location 5, the corrections show notable differences from the ASOS data, most likely due to its distance from an ASOS station. 

\begin{figure}
    \centering
\includegraphics[width=\textwidth]{   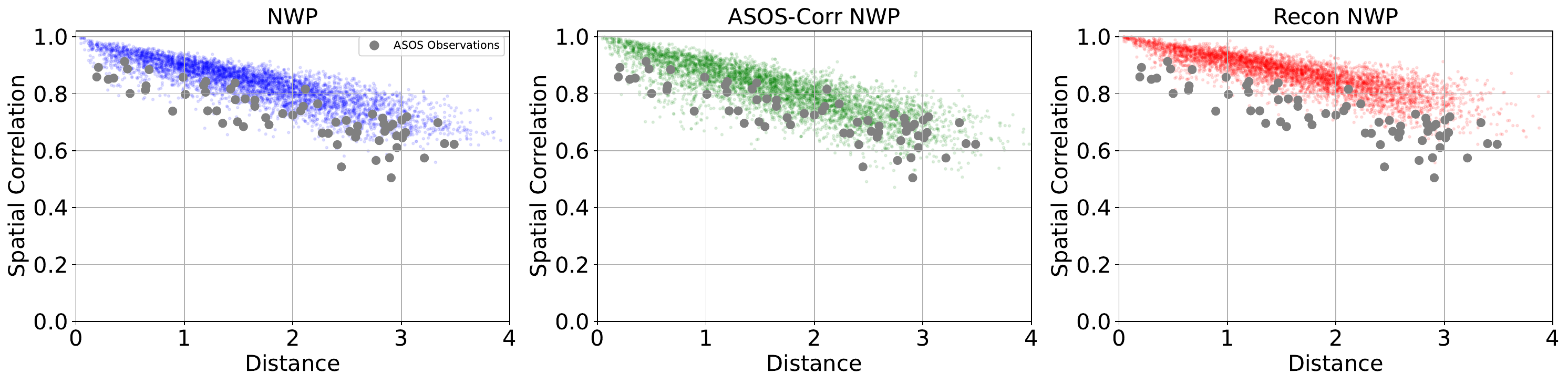}
    \caption{Correlation as a function of the distance between gridpoints. Grey circles represent the correlation of the original ASOS observation data. Blue points correspond to the original NWP data, green points represent the ASOS-corrected reconstructed NWP, and red points indicate the reconstructed NWP.}
    \label{fig:nwpcorr}
\end{figure}

Figure~\ref{fig:nwpcorr} shows the spatial correlation as a function of the distance between gridpoints. The reconstructed NWP (red) maintains the same general structure as the original NWP (blue) but exhibits slightly higher values, indicating that the reconstruction process may enhance or amplify certain spatial patterns. Both these datasets exhibit higher spatial correlation than the ASOS data (grey). 
However, the ASOS-corrected NWP (green) exhibit correlations closer to the observed ASOS ones, and more dispersed correlations compared to the original and reconstructed NWP. 

\begin{figure}
    \centerline{\includegraphics[width=0.75\textwidth]{   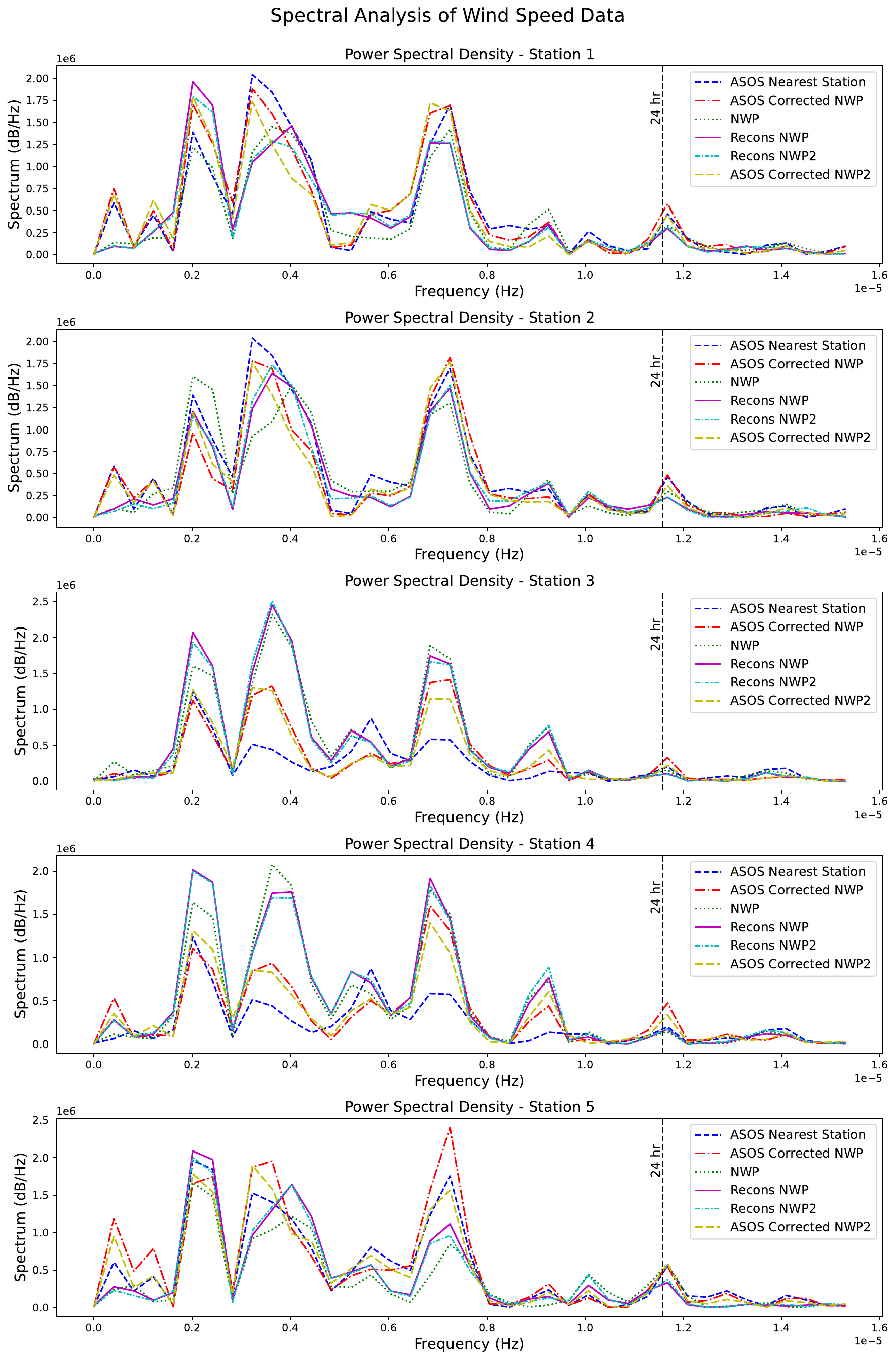}}
    \caption{Spectral analysis of the five selected stations across rows. The blue dotted lines represent the ASOS at the nearest observation station. The red line corresponds to the ASOS-corrected NWP using depth-1 Sig-PCA, the yellow line shows the ASOS-corrected NWP using depth-2 Sig-PCA, the green line represents the original NWP, the purple line denotes the reconstructed NWP using depth-1 Sig-PCA,  and the blue line indicates the reconstructed NWP using depth-2 Sig-PCA.}
    \label{fig:nwpspec}
\end{figure}

Figure~\ref{fig:nwpspec} shows the spectral analysis of the different method outputs at the five selected locations. Across all stations, the 24-hour peak in the wind spectrum is much smaller than the peaks at lower frequencies, (such as 6 days, 3 days and 1.7 days). This suggests that wind conditions are primarily influenced by multi-day atmospheric processes, such as weather systems, rather than by daily cycles or other short-term fluctuations. Overall, at most stations, depth-2 Sig-PCA (yellow) is better at capturing low-frequency variations than depth-1 Sig-PCA (red). The discrepancies between, respectively, the reconstructions and corrections with model data and observations are more significant with the wind speed dataset than the temperature application. This is mostly due to the high-variability of surface wind processes in space and time that challenges the reconstructive and corrective NN models as well as the summary statistics.  

\section{Discussion and Conclusion}\label{sec:conclu}

The proposed Sig-PCA approach integrates local observations with path signatures as summary statistics from large model outputs, significantly correcting and improving the accuracy of model outputs while working a reduced representation of these latter.  
The proposed method has been tested on two different datasets (surface temperature and surface wind speed), each with very different spatiotemporal structures, different ratios of model to observation data and different type of model outputs. The method provides more faithful corrections of temperature data than the surface wind ones, likely due to its stronger and smoother spatiotemporal structures. Most statistics used in validation align more accurately with the observational ones when the Sig-PCA is used in reconstruction and observation-based correction setting, highlighting the need to integrate both physics-based model outputs and measurements for most accurate and faithful representations. Finally, path signatures offer great flexibility as they capture spatiotemporal non-stationarity in the temperature data, as well as spatial heterogeneity in the wind speed data, lowering the need for covariates or non-stationary treatment, which remains typical statistical challenges. Path signatures also allow for higher-order summary statistics to capture more complex fields such as with surface wind speed. 
Finally, the use of local deep-kriging basis functions enhances the capture of spatial non-stationarity, which are most prevalent in the surface wind data. 

Our approach provides a significant reduction of the model outputs needed to train reconstructive and corrective NN. For temperature reconstruction, training on $4\%$ of the total gridded data and the proposed summary statistics yields reconstructed model outputs with an average of $11.05\%$ error across the entire grid. When observations are incorporated, the $\%$ improvement error is 70.09$\%.$  For the wind NWP reconstruction with $4\%$ model data, an error of $13.63\%$ is obtained to reconstruct NWP outputs only, while with the correction NN, we get a $\% $ improvement error of 51.61$\%$ with respect to the model outputs, highlighting the increased complexity of correcting the surface winds. 

While our method allows a reduction in training data and reliable subsequent corrections, it relies on evaluating the summary statistics at every gridpoint of the model outputs.  
A first attempt was made to compute path signature at the $k$-means centroids on each considered model output. 
The reconstruction error of NLDAS temperature averages around $25.75\%$, significantly worsening the reconstruction quality that only carried $11.05\%$ of error when SigPCA is used. 
Meanwhile, the reconstruction of NWP wind based on $k$-centroids lead to $14.24\%$ error, nearing the reconstruction error when SigPCA is employed. 
We suspect that the smoothness of temperature fields lead to a poorly meaningful clustering while wind fields, which typically exhibit spatial heterogeneity, provided a more informative clustering. 
This highlights the strength of path signatures in capturing most relevant information to reconstruct model outputs; while emphasizing the need for sophisticated spatial sampling techniques when computing path signature at every gridpoint becomes computationally expensive. 
Another approach could rely on adaptive sampling to identify the optimal data points to represent the spatial grid; however, current methods for spatial adaptive sampling are application dependent \citep{ROUETLEDUC20141857} and general methods remain under-developed.  
Finally, incorporating space-time interaction into path signatures may be an additional modeling step; however, it would still require subseting relevant spatial gridpoints to lower computational costs.

The proposed framework has strong potentials for in-situ implementation of dimension reduction of large model output data, ensuring that key statistical features are extracted at the source, rather than requiring the movement of massive datasets. 
Recent advancements have been made in the context of in-situ techniques and implementations, for instance where distribution-based summaries are developed for multivariate data  \citep{hazarika+dsc19}, representative temporal in-situ summaries are proposed in \citet{dutta+etal21} without any reconstruction focus, and a Julia-based framework for in-situ data analysis \citep{tang2024julia}.

\section*{Acknowledgments}
We thank Emil Constantinescu for sharing the NWP wind speed data from earlier study \citep{bessac+ca18}. We thank Wenqi (Flora) Zhang for helpful discussions about measurement data. We thank Johann Rudi for their discussion and advice. 
This work was authored by the National Renewable Energy Laboratory, operated by Alliance for Sustainable Energy, LLC, for the U.S. Department of Energy (DOE) under Contract No. DE-AC36-08GO28308. Efforts are funded by the DOE Office of Science Early Career Research Program. 
The views expressed in this article do not necessarily represent the views of the
DOE or the U.S. Government. The U.S. Government retains and the publisher, by
accepting the article for publication, acknowledges that the U.S. Government retains
a nonexclusive, paid-up, irrevocable, worldwide license to publish or reproduce the
published form of this work, or allow others to do so, for U.S. Government purposes. 
The research was performed using computational resources sponsored by the Department of Energy's Office of Energy Efficiency and Renewable Energy and located at the National Renewable Energy Laboratory.

\section*{Code}
All codes for the two examples are available at \url{https://github.com/atlantac96/Sig-PCA.git}

\section{Appendix}

\section*{Additional Results for Temperature}

\begin{figure}
    \centering
\includegraphics[width=\textwidth]{   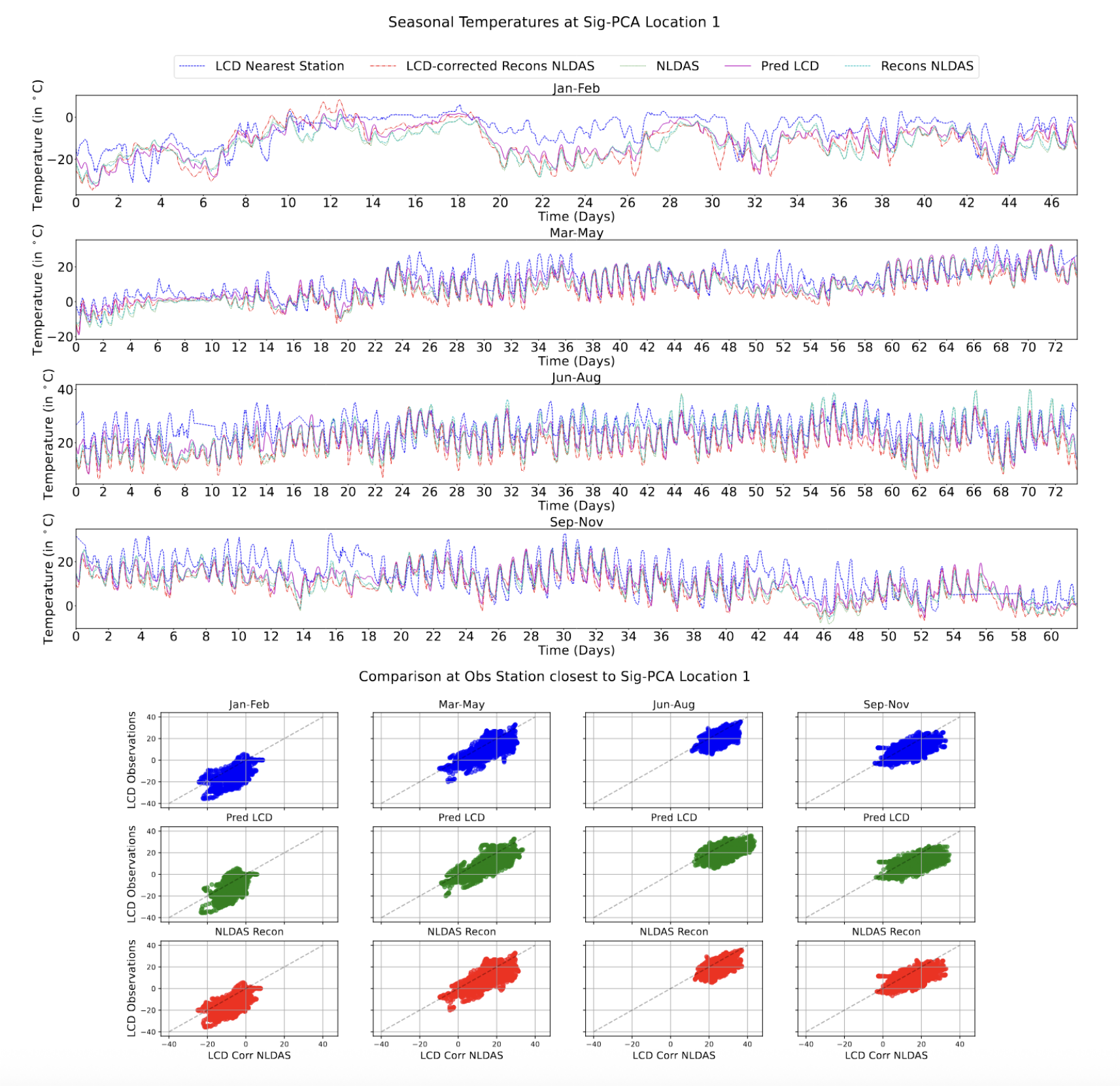}

    \caption{\textbf{Top:} Comparison of different time series  for Location 1 with the time series of nearest observation station. The LCD-corrected NLDAS is denoted in red, NLDAS in green, directly outputted LCD in purple and reconstructed NLDAS in cyan. \textbf{Bottom:} QQ-plot of outputs for different approaches (x-axis) compared with the LCD observations (y-axis) at the nearest observation station to Sig-PCA at Location 1. The rows represent predicted LCD (top in blue), reconstructed NLDAS (middle in green), and LCD-corrected reconstructed NLDAS (bottom in red), while the columns correspond to the four different seasons.}
    \label{fig:6loc1}
\end{figure}

\begin{figure}
    \centering
\includegraphics[width=\textwidth]{   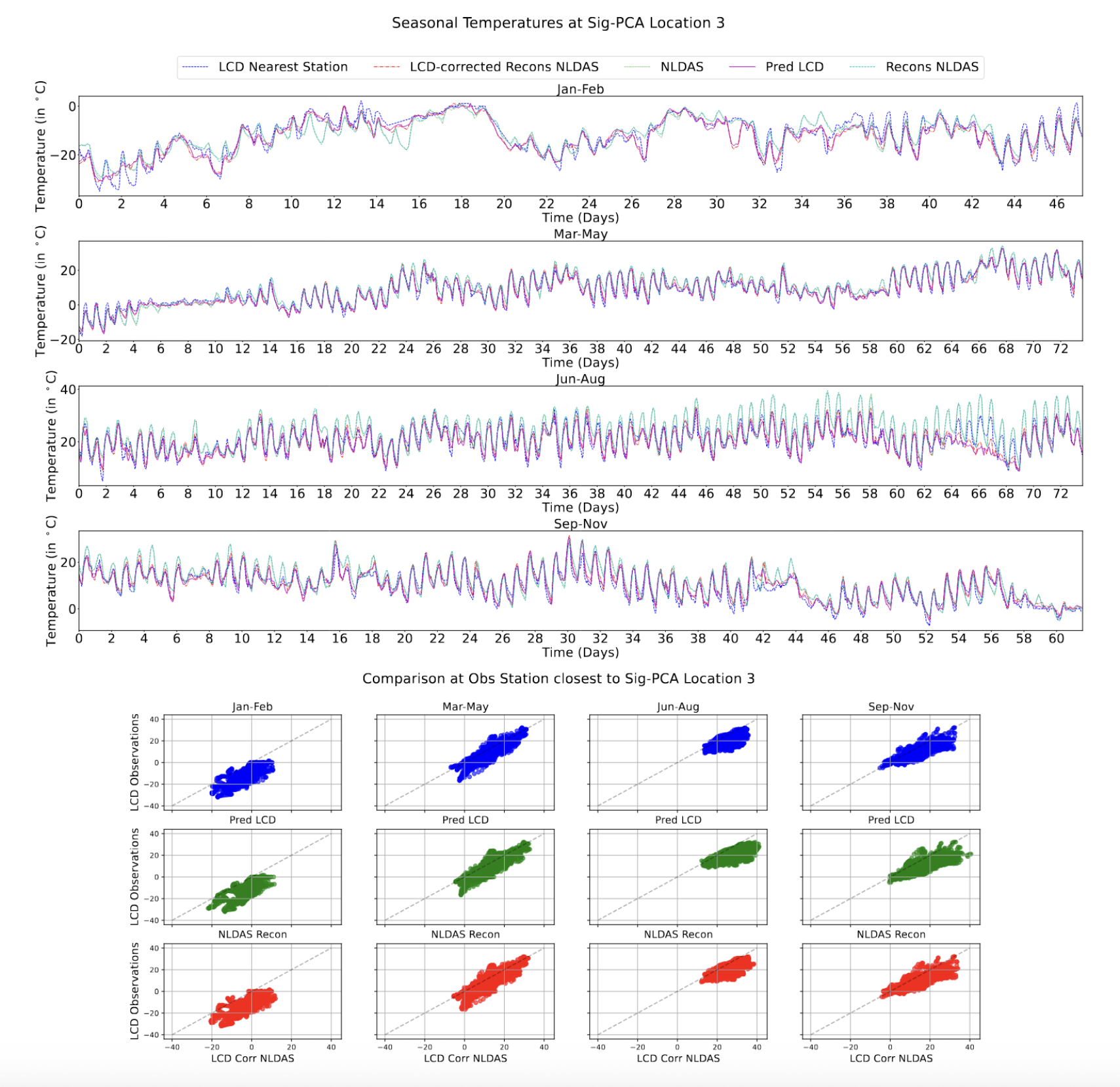}

    \caption{\textbf{Top:} Comparison of different time series  for Location 3 with the time series of nearest observation station. The LCD-corrected NLDAS is denoted in red, NLDAS in green, directly outputted LCD in purple and reconstructed NLDAS in cyan. \textbf{Bottom:} QQ-plot of outputs for different approaches (x-axis) compared with the LCD observations (y-axis) at the nearest observation station to Sig-PCA at Location 3. The rows represent predicted LCD (top in blue), reconstructed NLDAS (middle in green), and LCD-corrected reconstructed NLDAS (bottom in red), while the columns correspond to the four different seasons.}
    \label{fig:6loc3}
\end{figure}

\begin{figure}
    \centering
\includegraphics[width=\textwidth]{   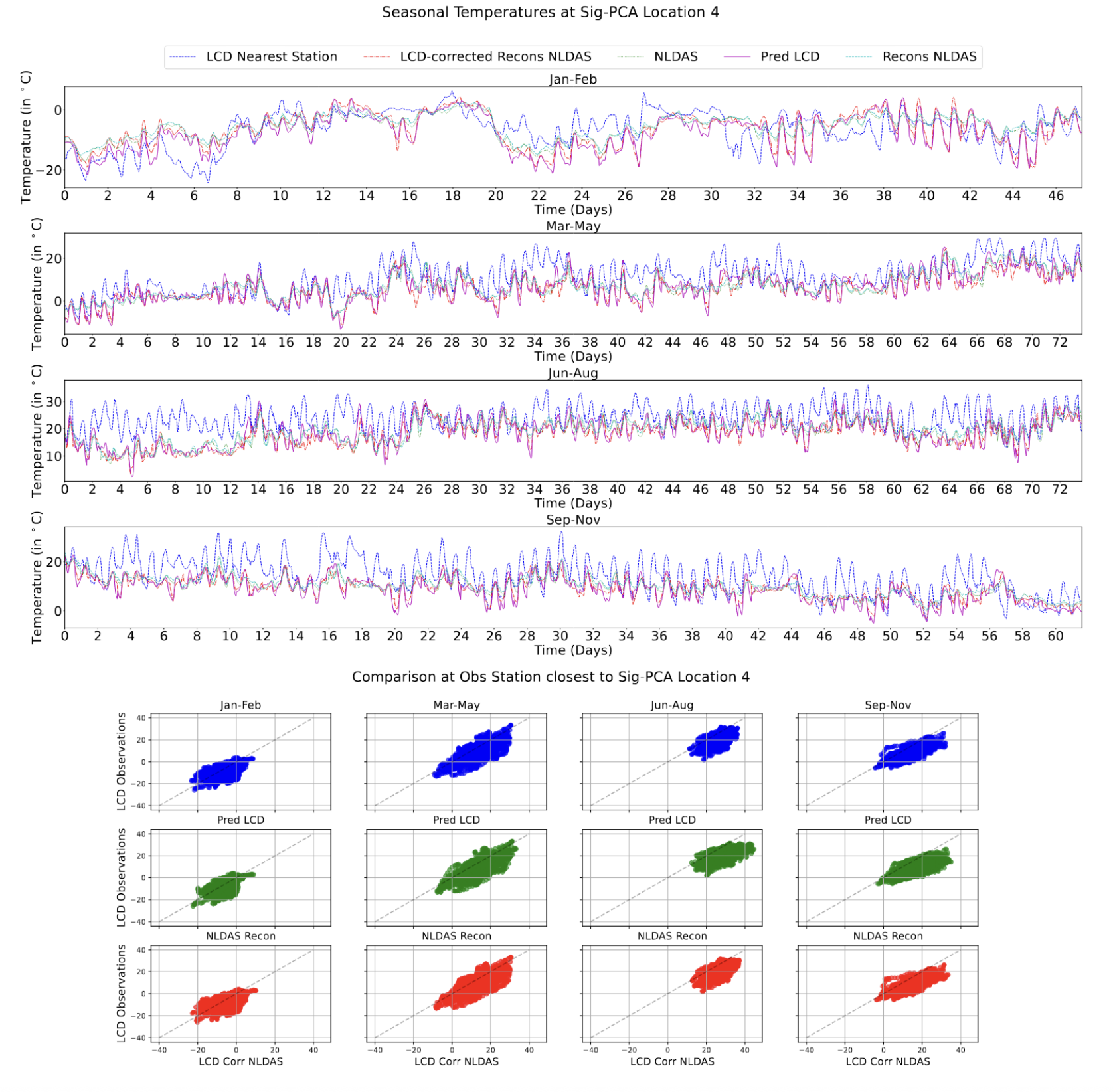}

    \caption{\textbf{Top:} Comparison of different time series  for Location 4 with the time series of nearest observation station. The LCD-corrected NLDAS is denoted in red, NLDAS in green, directly outputted LCD in purple and reconstructed NLDAS in cyan. \textbf{Bottom:} QQ-plot of outputs for different approaches (x-axis) compared with the LCD observations (y-axis) at the nearest observation station to Sig-PCA at Location 4. The rows represent predicted LCD (top in blue), reconstructed NLDAS (middle in green), and LCD-corrected reconstructed NLDAS (bottom in red), while the columns correspond to the four different seasons.}
    \label{fig:6loc4}
\end{figure}

\begin{figure}
    \centering
 \includegraphics[width=\textwidth]{   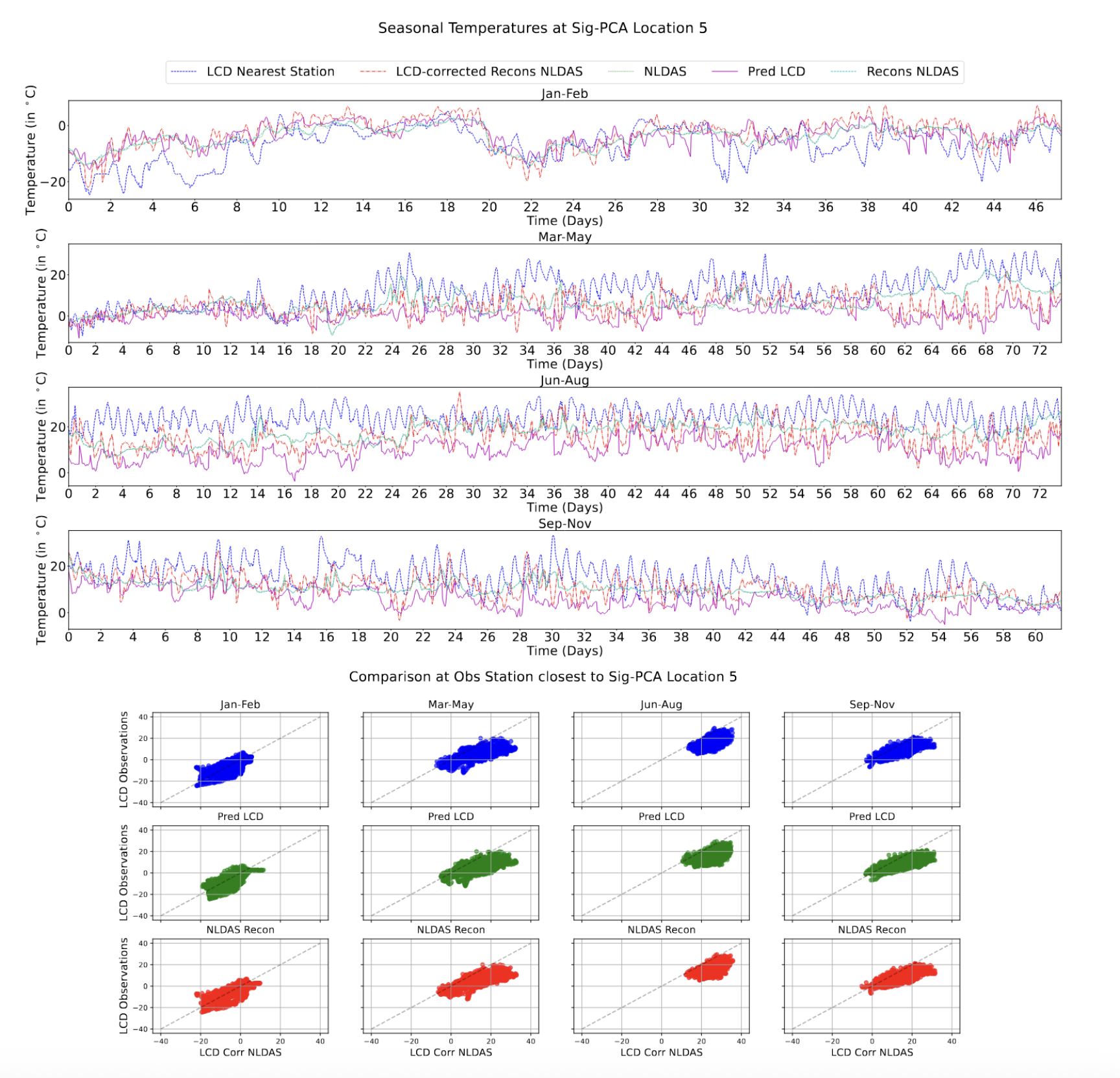}
\caption{\textbf{Top:} Comparison of different time series  for Location 5 with the time series of nearest observation station. The LCD-corrected NLDAS is denoted in red, NLDAS in green, directly outputted LCD in purple and reconstructed NLDAS in cyan. \textbf{Bottom:} QQ-plot of outputs for different approaches (x-axis) compared with the LCD observations (y-axis) at the nearest observation station to Sig-PCA at Location 5. The rows represent predicted LCD (top in blue), reconstructed NLDAS (middle in green), and LCD-corrected reconstructed NLDAS (bottom in red), while the columns correspond to the four different seasons.}
    \label{fig:6loc5}
\end{figure}

Top panel of \Cref{fig:6loc1,fig:6loc3,fig:6loc4,fig:6loc5} shows the predicted time series across locations 1, 3, 4 and 5, whereas the bottom panel shows the QQ plots comparing the outputs of different methods to the nearest observation station for the highlighted locations. Since these observation stations are located far from the Sig-PCA locations, the outputted observations are more spatially dispersed and do not directly match the LCD observations at the nearest station.

Across all locations, the reconstruction aligns closely with the original NLDAS data. At Location 1, the largest discrepancy between the LCD at the nearest station and the NLDAS occurs during winter, while in spring, both the corrections and the direct LCD output are similar. In summer and fall, the corrections align more closely with the NLDAS, especially since the observational station is farther away. At Location 3, the direct LCD output and the corrections are nearly identical and both mimic the observations well across all seasons. Additionally, the model outputs consistently show higher values than the LCD fine-scale observations, highlighting the discrepancy between the two datasets. For Location 4, a noticeable discrepancy between NLDAS and the LCD at the nearest station is most pronounced in winter. The corrections and direct LCD output are similar, but neither can accurately match the LCD observations due to the distance from the nearest observation station. Finally, at Location 5, the direct LCD output and corrections are quite different. The corrections are able to capture the peaks of the observational data, which the direct LCD output cannot. This trend holds across all seasons, with the corrections providing a better match to the observations than the direct LCD output. Location 5 is the farthest from any observation station compared to the other highlighted locations.

\begin{figure}[H]
    \centering
\includegraphics[width=0.8\textwidth]{   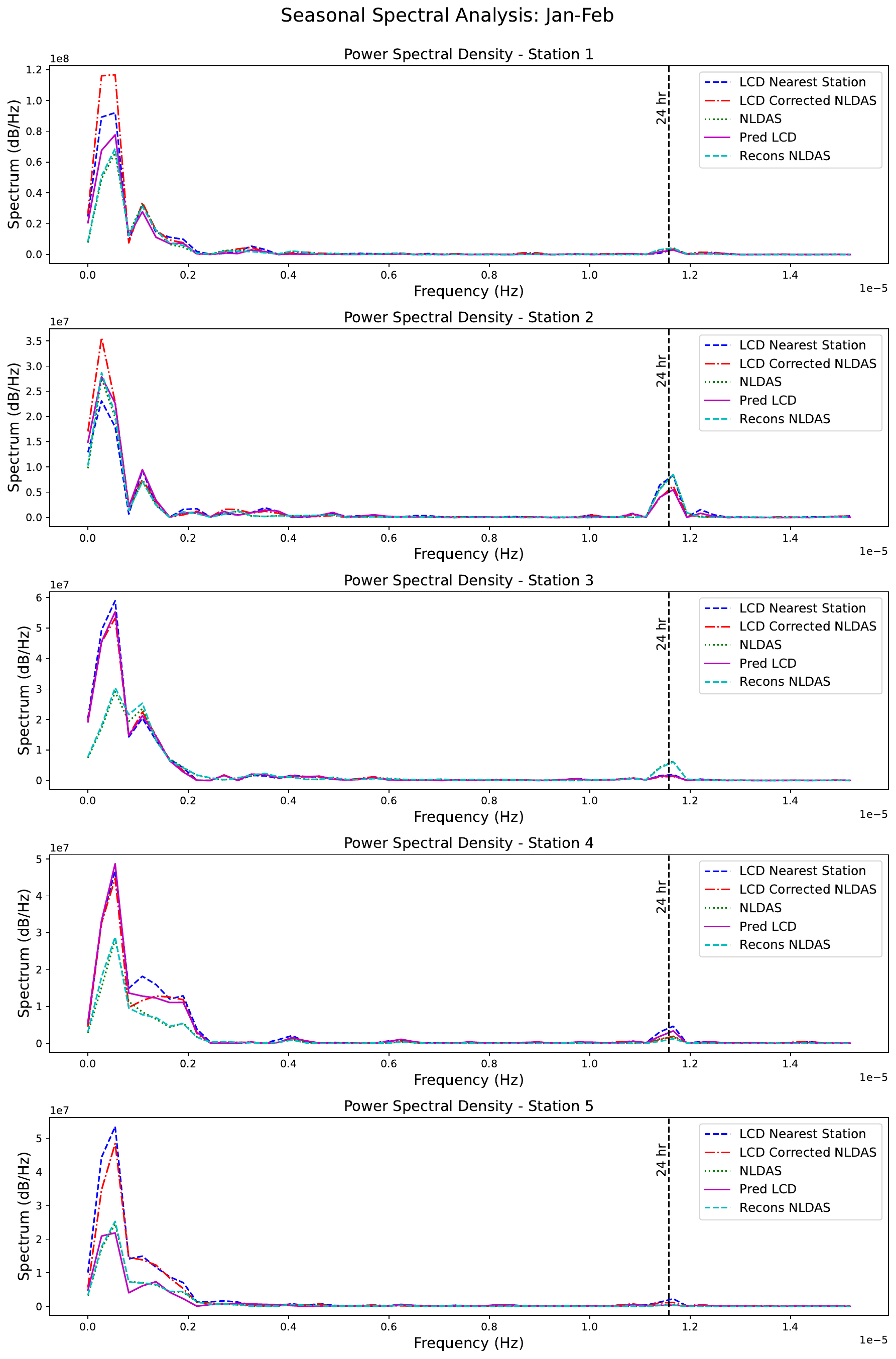}
    \caption{Seasonal spectrum for winter (Jan-Feb)}
    \label{fig:janfebspectra}
\end{figure}

\begin{figure}[H]
    \centering
\includegraphics[width=0.8\textwidth]{   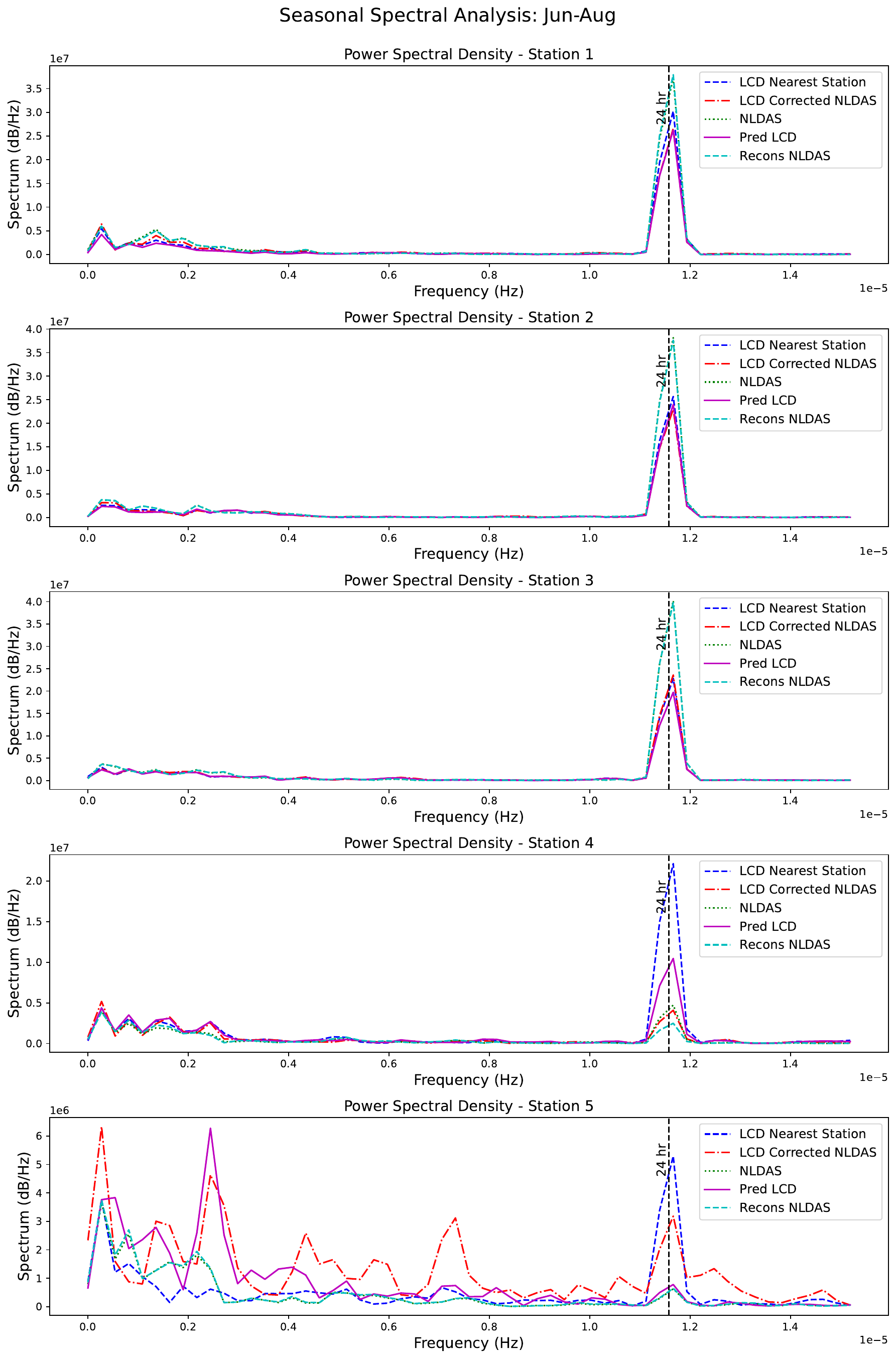}
    \caption{Seasonal spectrum for summer (June-Aug)}
    \label{fig:juneaugspectra}
\end{figure}

\begin{figure}[H]
    \centering
\includegraphics[width=0.8\textwidth]{   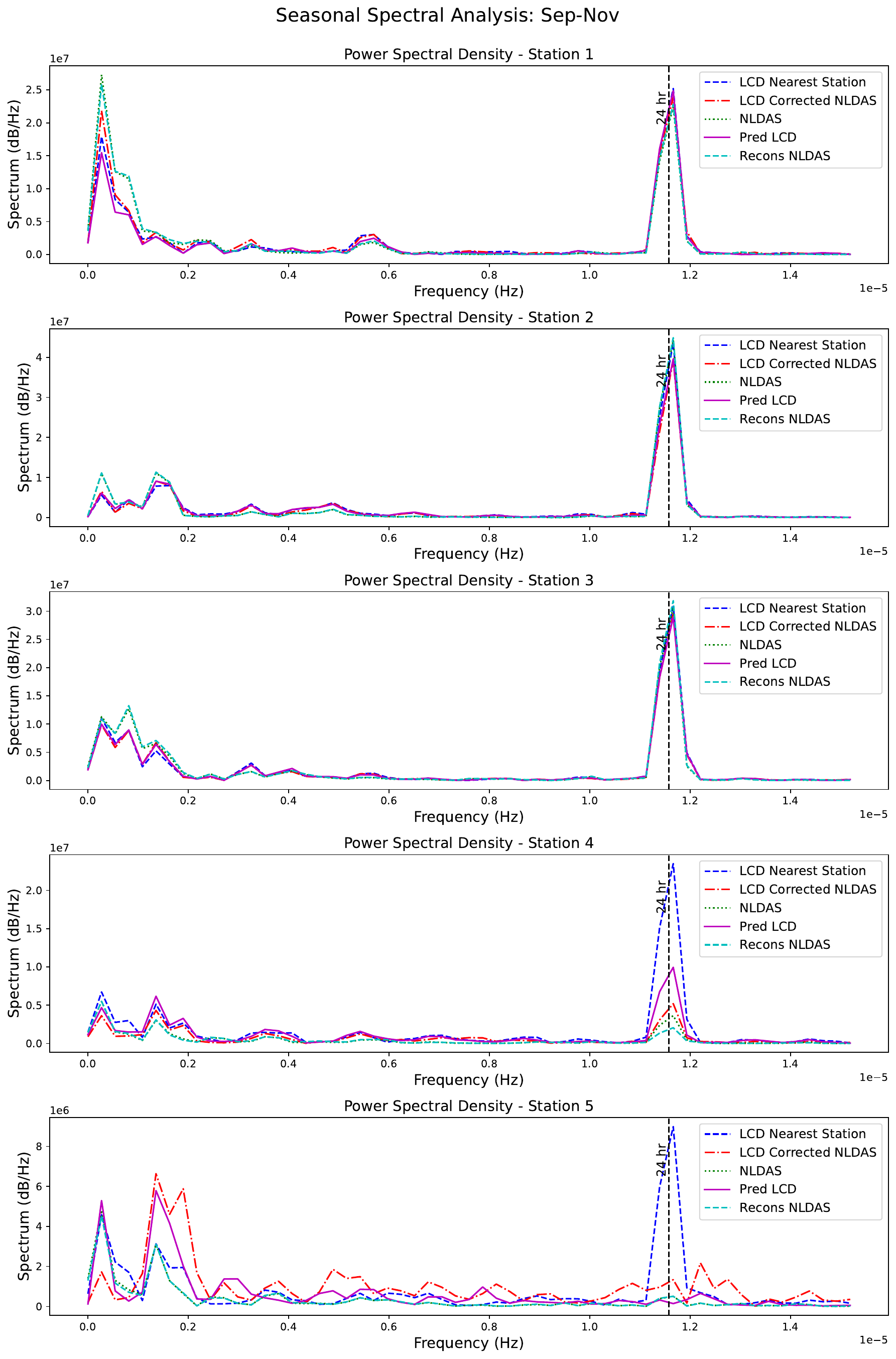}
    \caption{Seasonal spectrum for winter (Sep-Nov)}
    \label{fig:sepnovspectra}
\end{figure}

\Cref{fig:janfebspectra,fig:juneaugspectra,fig:sepnovspectra} show the seasonal spectrum across the 5 highlighted locations. In winter, at location 5, the direct LCD output closely matches NLDAS at lower frequencies, while the LCD correction aligns better with the nearest observation station, capturing localized variability. This suggests that the direct LCD output is largely influenced by the coarse NLDAS data at lower frequencies, meaning it retains the broad-scale patterns without substantial modification. However, the LCD-correction method incorporates observational data, allowing it to align more with the local station rather than the broader NLDAS trends. For locations 3 and 4, both the direct LCD output and LCD correction match the lower-frequency peaks of the nearest station instead of NLDAS, reflecting the influence of local observations. Unlike location 5, here both methods deviate from NLDAS at lower frequencies, implying that the local observational data have a stronger impact on the final reconstruction. This could be due to more significant discrepancies between NLDAS and local climate conditions in these areas as can be seen in \Cref{fig:6loc3,fig:6loc4}, making the local LCD data more dominant in shaping the spectral features.
However, at locations 1 and 2, the LCD correction exhibits a stronger peak at lower frequencies than the nearest station, while the direct LCD output remains more aligned with NLDAS at the 24-hour mark. The LCD correction introduces additional low-frequency components, possibly due to over-correction or amplification of certain localized patterns. In contrast, the direct LCD output continues to be primarily influenced by NLDAS, maintaining the dominant daily cycle structure seen in the original dataset.  

In summer, the 24-hour peak remains stronger than lower-frequency components for all stations except station 5, where LCD-corrected NLDAS introduces several additional peaks compared to the direct LCD output, which instead mimics the nearest station's LCD data. Summer is typically characterized by strong diurnal heating cycles, which explains why the 24-hour peak dominates. However, at station 5, the LCD correction method introduces additional variability at lower frequencies, potentially due to atmospheric processes that are captured better through local observations rather than coarse-scale NLDAS data at that particular location. 
In fall, Location 1 uniquely exhibits a strong lower-frequency peak, though the 24-hour cycle remains the most pronounced across all locations, reinforcing the regulairty of diurnal variations. The presence of a strong lower-frequency peak at Location 1 suggests long-term variations, possibly due to transitional weather patterns in fall, such as prolonged periods of warm or cool temperatures. Station 5 again shows multiple peaks in the lower frequencies. The presence of multiple peaks at lower frequencies suggests that station 5 experiences long-term variability possibly driven by topographic effects, or regional climate influences. 

\bibliographystyle{chicago}
\bibliography{oldreferences}

\end{document}